\documentclass[journal]{IEEEtran}
\pdfoutput=1
\usepackage{blindtext}
\usepackage{graphicx}
\usepackage{subfigure}
\usepackage{algorithm}
\usepackage{pseudocode}
\usepackage{algpseudocode}
\usepackage{amsmath}

\usepackage[hyphens]{url}
\usepackage{endnotes}
\usepackage{commath}
\usepackage{mathtools}

\ifCLASSINFOpdf
\else
\fi
\begin{document}\sloppy
%
\title{Accurate Online Full Charge Capacity Modeling of Smartphone Batteries}
%
%
%

\author{Mohammad~A. Hoque,
        Matti~Siekkinen,
        Jonghoe Koo,
        and~Sasu~Tarkoma
\thanks{Mohammad A. Hoque and Sasu Tarkoma are with the Department of 
Computer Science, University of Helsinki, Finland. Email: firstname.lastname@cs.helsinki.fi}
\thanks{Matti Siekkinen is with the Department of Computer Science and Engineering, Aalto University, Finland. Email: matti.siekkinen@aalto.fi}
\thanks{Jonghoe Koo is with the Department of Electrical and Computer Engineering and INMC, Seoul National University, Seoul, Korea. Email: jhkoo@mwnl.snu.ac.kr}
}

\maketitle

\begin{abstract}

Full charge capacity (FCC) refers to the amount of energy a battery can hold. It is the fundamental property of smartphone batteries that diminishes as the battery ages and is charged/discharged. We investigate the behavior of smartphone batteries while charging and demonstrate that the battery voltage and charging rate information can together characterize the FCC of a battery. We propose a new method for accurately estimating FCC without exposing low-level system details or introducing new hardware or system modules. We also propose  and implement a collaborative FCC estimation technique that builds on crowdsourced battery data. The method finds the reference voltage curve and charging rate of a particular smartphone model from the data and then compares the curve and rate of an individual user with the model reference curve. After analyzing a large data set, we report that 55\% of all devices and at least one device in 330 out of 357 unique device models  lost some of their FCC. For some models, the median capacity loss exceeded 20\% with the inter-quartile range being over 20 pp.  The models enable debugging the performance of smartphone batteries, more accurate power modeling, and energy-aware system or application optimization. 
\end{abstract}


%
\IEEEpeerreviewmaketitle


\section{Introduction}

Smartphone users frequently encounter battery and energy problems. 
From the popular Internet blogs~\cite{nexus6,iphone5,andforum}, we have identified two issues that are increasingly being reported by the users; sudden drop in the battery level and disgraceful shutdown of the device even with high battery levels being reported to the user (even at 80\%) while discharging. These observations are reported across different smartphone models, and even for laptops. This disgraceful shutdown may bar users from their scheduled phone activities  and result in data loss. From a user's perspective, the remaining battery life of a smartphone may even converge to the monetary value~\cite{Hosio:2016}. Many smartphone manufacturers have recently introduced battery replacement programs that cover batteries that have a reduced capacity, typically below 80\%~\cite{breplacement1,applebat}. The current smartphone battery discussion pertains to the following questions: {\emph{Why does the battery level fluctuate? Is the battery faulty? Is the problem due to an operating system upgrade or installing/upgrading an application?}}.

Our prior work~\cite{mhotpower2015} demonstrated that the answers to the earlier questions are related to the smartphone full charge capacity (FCC).   FCC is the maximum amount of charge an empty battery can hold. As the battery of a smartphone ages, the full charge capacity decreases with the utilization. FCC is typically modeled as a function of the number of charging cycles or the age of the battery, or measured with Coulomb counting technique. This functionality or measurement capability resides inside a smart battery. The battery shares this estimates as percentage with the hosting device, such as smartphone (see Section~\ref{sec:soc}). Therefore, an indication of the capacity loss along with the battery level would allow more accurate power consumption modeling, sophisticated energy-aware scheduling mechanisms by the system and  different  applications.

In this article, we examine the performance of smartphone batteries and present a novel FCC estimation technique that can infer the FCC and FCC loss. The approach works whether a smart battery is capable to measure FCC or not, and therefore enables any device to estimate the FCC of the attached battery. We discover that the battery voltage and battery capacity relative charging rate, i.e., C-rate, curves can characterize the FCC of a smartphone battery given the reference curves of the new battery. Based on these findings, we devise a new FCC estimation method.  Our evaluation suggests that the estimation error is limited to 10\% of the true value. Our technique can be implemented, for instance, as a mobile application or it can be integrated into the operating system of the mobile device to monitor battery capacity health without the need to deploy new hardware. To the best of our knowledge, our FCC estimation technique is a new approach.

In order to facilitate large scale battery health analytics, we also present a crowdsourced approach that works with battery voltage and charging rate information solely obtained from a crowdsourced data set. In order to study the capacity loss of the devices that contribute to such a data set, we derive a reference voltage and a C-rate curve from the data set for each model using a statistical approach, and then apply our FCC estimation method that compares the charging rate of a device with the model specific reference rate. We demonstrate that this method works relatively well for most models found in the Carat data set~\cite{Oliner2013}. Furthermore, we discovered that 55\% of all the 9560 devices had some capacity loss, and that within 357 unique models at least one device in 330 models suffered from capacity loss. We summarize our contributions as follows.


\begin{itemize}

\item We investigate the behavior of battery voltage of smartphones while charging and reveal the relationship between battery voltage and the remaining battery capacity. The voltage for a specific state of charge (SOC) of a battery with reduced capacity is higher than that of a new battery, and the voltage reaches its maximum value at a lower value of the SOC as the FCC decreases.   We also investigate the charging rate behavior and observe that the relative charging rate of the battery increases as the FCC decreases. Consequently, we propose and validate a new battery full charge capacity estimation model, which works with different SOC estimation techniques. The model yields estimates with an accuracy of 90\% or more according to our evaluation. 

\item We introduce a crowdsourced FCC estimation technique. Although the accuracy of the approach depends on the diversity of the community, a study with the large scale Carat data set shows that our collaborative technique estimates the reference rates of popular smartphone models within a 10\% error margin. We also examine the presence of devices with reduced FCC in the Carat data set. 
\end{itemize}

The paper is organized as follows: Section~\ref{sec:soc} presents a smartphone power management primer. In Section~\ref{sec:three}, we investigate battery voltage and charging current behavior while charging the smartphones. We present and validate a rate-based FCC estimation model in Section~\ref{sec:fcc}. The crowdsourced battery analytic data set and the battery FCC estimation technique are described in Section~\ref{sec:five}.  The limitations of our FCC estimation methods are discussed in Section~\ref{sec:seven} and compared with related work  in Section~\ref{sec:eight}. Finally, we draw conclusions from this paper in Section~\ref{sec:nine}.

\section{Smartphone Power Management}
\label{sec:soc}
\begin{figure}[t]
 \begin{center}
 \includegraphics[width=0.8\linewidth,height = 0.7\linewidth]{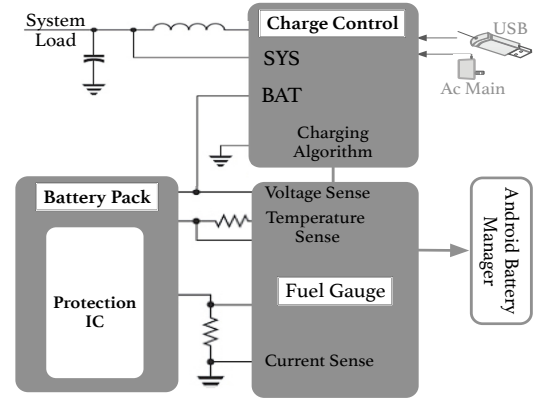}
   \caption{{\bf Smartphone Power Management Hardware Block Diagram } {\sl The charging controller divides the current drawn from the charger for the battery based on system load. The battery receives less charging current if the device is being used while charging.}}
    \label{fig:charging_block}
 \end{center}
\end{figure}

Smartphone power management comprises a battery, a fuel gauge chip, and a charging controller. The chip with SOC estimation functionality is often called fuel/gas gauge. The functionality of a fuel gauge chip may be distributed between the battery pack and the host system, i.e., smartphone~\cite{fuelsel}. The smartphone only queries the battery for the supported information, such as SOC, battery voltage, and temperature. The block diagram in Figure~\ref{fig:charging_block} illustrates how these components work together.

\subsection{Charging Controller}
Smartphones are mostly charged using either an AC wall or a USB charger. The charging controller applies Constant Current-Constant Voltage (CC-CV) charging method. During the CC period, the charging current is constant until the battery voltage reaches a specified maximum (4.2/4.35V), after which the charging current is trickled until the battery is fully charged. The charging terminates when the charging rate reduces to 0.07C or to a lower cut-off charging current specified by the manufacturer~\cite{Thanh:2012}. In this case, C is the rate that is relative to the battery capacity as follows: If the capacity of a battery is 2600 mAh and it takes one hour to fully charge/discharge a battery, it means that 2600 mA rate is equivalent to 1 C for that battery. Similarly, 0.5 C-rate is equivalent to 1300 mA for that battery. In presence of system load, the charging controller may deliver lower current to the battery pack (see Section~\ref{sub:chg_proc}).

\subsection{Battery Pack}
Smartphones are powered with single cell battery packs. Along with the cell, a battery pack may also host SOC/FCC measurement functionalities and such batteries are called smart batteries. The battery pack may also include a protection mechanism, as shown in Figure~\ref{fig:charging_block}, to guard against higher voltage and current from the device. The FCC of a battery decreases as it ages and it is an irreversible process. The capacity reduction happens through progressive chemical reactions. Graphite is the common material used as anode in Lithium-Ion batteries and there are multiple anode-cathode pairs in a battery. As the battery is being charged, the oxidization of the graphite constructs a layer, called passive surface layer~\cite{Kida20024157}. 
If the outer shell is leaked, then the oxidization happens faster due to moisture and the capacity loss accelerates.

\subsection{Fuel Gauge}
Based on the manufacturer, a fuel gauge chip may be able to measure or estimate the FCC along with SOC. 

\noindent\textbf{SOC estimation}~~The SOC is the runtime estimate of the battery charge. A SOC value of 0 and 100 imply an empty and fully charged battery, respectively.  The most common approach to estimate SOC is to use open circuit voltage (OCV) with a number of look-up tables. A voltage based fuel gauge may also combine both OCV and load voltage to estimate SOC or energy drain~\cite{Koo:2016, maxim17048}. The second approach is Coulomb counting, which introduces a sense register on the charge and discharge path as shown in Figure~\ref{fig:charging_block}. Table~\ref{tab:fuelgauge_info} describes a number of SOC estimation techniques used by the modern fuel gauge chips. 

\begin{table}[t]
\begin{center}
  \caption{Characteristics of different Fuel Gauge Solutions. Coulomb counter-based solutions mostly reside in the battery.}

  {\footnotesize
    \begin{tabular}{|p{12mm}|p{28mm}|p{35mm}|}
      \hline      
       SOC~FCC\break Solution&Advantage&Disadvantage\\\hline
OCV Lookup Table & 
(1) Only voltage measurements,
(2) A number of look-up tables& 
(1) Increasing SOC error with dynamic system load,
(2) Cannot report FCC\\\hline
Coulomb Counter & 

(1) Report SOC(\%) and remaining capacity (mAh),
(2) Accurate SOC for single discharge&
(1) Need full charge and discharge learning cycle,
(2) Need current sense resistor\\\hline

Voltage Dynamic Model&
(1) Eliminate learning cycle,
(2) Report SOC (\%)
 &
(1) Slow response time for dynamic load,
(2) Cannot report remaining capacity(mAh) due to no information of current and full capacity\\\hline
  \end{tabular}}
         \label{tab:fuelgauge_info}
    \end{center}
 
\end{table}




\noindent\textbf{FCC estimation}~~ Table~\ref{tab:fuelgauge_info} shows that voltage based approaches cannot estimate or measure the FCC. Therefore, a fuel gauge chip may use the number of charging cycles that the battery has gone through to estimate the FCC. A charging cycle is equivalent to a complete discharge of a battery from a full to an empty state. One charging cycle can comprise multiple discharge events. A fuel gauge may use FCC learning by mapping FCC with charging cycles, temperature, and OCV.   Coulomb counter-based fuel gauges use sense resistors to measure FCC. However, the measurements can be used internally by the fuel gauge to recalibrate the SOC and may not be shared with the smartphone. Such functionalities must reside inside the battery pack as the battery may be changed.

This article proposes and validates a new software-based FCC estimation technique. Unlike the above approaches, \textit{our approach neither requires complex learning nor additional hardware. The method equally works with the devices powered by both OCV and Coulomb counting-based fuel gauges}.

\section{Smartphone Charging Behavior}
\label{sec:three}

In this section, we first investigate the charging behavior of smartphones with different charging configurations and system load. We next investigate the performance of three different types of Android smartphone batteries, namely new batteries, new batteries with lower capacity, which we call \textit{substandard} batteries, and long used, aged batteries whose capacity has reduced. The aim of the experiments is to understand how the battery voltage behavior changes as the capacity of a battery decreases and to use the lessons learned to derive a method to estimate FCC of a smartphone battery. 

\subsection{Experiment Setup}
We investigate the performance of Samsung Galaxy S2 (GT-I9100), S3 (GT-I9300), and S4 (GT-I9505) batteries while \textit{charging}. The reason for selecting these devices is that their batteries are replaceable. We used total 15 batteries of 1650, 2100, and 2600 mAh, and a substandard battery of 2000 mAh from a third party battery manfacturer. Compared to our earlier work~\cite{mhotpower2015}, we included Galaxy S3 and ten additional batteries in this study.  Among these three devices, Galaxy S3 is equipped with a fuel gauge which uses both OCV and load voltage to estimate SOC. The others use simple OCV-based fuel gauges.

\begin{figure}[t]
  \begin{center}
   \includegraphics[width=0.7\linewidth,height = 0.3\linewidth]{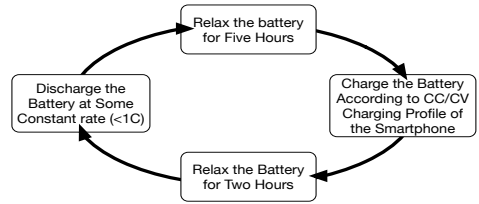}
    \caption{The smartphone charging and discharging  procedure followed during the exepriments.}
    \label{fig:chargingexperiment}
 \end{center}
 \end{figure}

The measurements  are divided in two sets. In the first  set, we experimented the charging controller behavior with different charging configurations and system load (see Section~\ref{sub:chg_proc}). In the  second set of  charging measurements, we used USB2.0 and AC wall charger, and kept the smartphones idle in airplane mode in order to maintain a constant current supply to the battery pack. At the beginning of each charging measurement, we discharged the battery by keeping the display ON with a fixed brightness level, then relaxed the battery for five hours, and finally charged the device as illustrated in Figure \ref{fig:chargingexperiment}. The low rate discharge ensures an empty battery. The fuel gauge manufacturers also conduct their experiments in this way \cite{impedence}. During the measurements, the room temperature was 21-25$^{\circ}$ Celsius and each experiment was repeated four times.


We collect and analyze the battery analytics data from the devices. The BatteryManager in Android devices collects battery voltage, health, and temperature from the fuel gauge as shown in Figure~\ref{fig:charging_block} and broadcasts such information whenever there is a change in the battery level or a charger is plugged/unplugged. We modified the Android version of Carat application~\cite{Oliner2013} to store only the battery API provided information in the smartphone.

\begin{figure}[t]
  \begin{center}
  \includegraphics[width=0.8\linewidth,height = 0.53\linewidth]{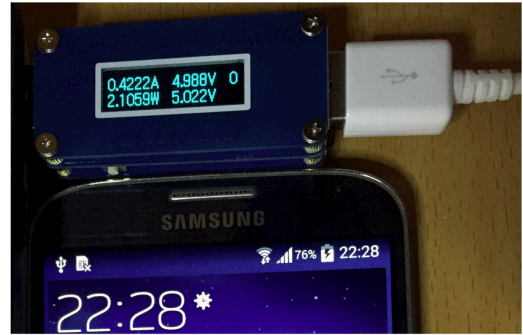}
  
    \caption{{\bf Charging current and the FCC measurement with Yoshoo power monitor. }{\sl Yoshoo is a simple USB3.0 pass-through power measurement tool.}}
    \label{fig:soc_measure}
 \end{center}
 \end{figure}

\begin{table}[t]
\begin{center}
  \caption{Smartphones' charging current behavior during the CC phase with different chargers while the devices are turned OFF.}

    \begin{tabular}{|p{15mm}|p{12mm}|p{15mm}|p{12mm}|p{10mm}|}
      \hline      
       Model&Stock AC Charger&Alternate Charger& Charging\break Current & Recognized Charger\\\hline
       Galaxy S4& 5V, 2.1A& 5V, 1.0A, AC& $\approx$0.930A& AC \\\hline
       Galaxy S3 & 5V, 1A & 5V, 2.1A, AC &$\approx$0.930A& AC\\\hline
       Galaxy S2 & 5V, 0.7A & 5V, 2.1A, AC &$\approx$ 0.650A& AC\\\hline
       Galaxy S2 & 5V, 0.7A & 5V, 2.0A, Battery Pack & $\approx$ 0.650A& AC\\\hline

  \end{tabular}
         \label{tab:rates_cap}
    \end{center}
 
\end{table}

\subsection{Demystifying Smartphone Charging Procedure}
\label{sub:chg_proc}
Whenever a device is connected to the charger, the charging controller first checks the OCV and SOC to determine the charging phase and next draws current accordingly from the charger. As we have investigated using a USB3.0 Yoshoo power monitor (see Figure~\ref{fig:soc_measure}) with USB2.0, AC wall chargers and an external battery pack, the maximum charging current, ($I_{chg}$) drawn by the charging controller can be defined as $min(charger~output,~controller~current)$ during the CC phase. For instance, the charging controller of Galaxy S2 draws a maximum of 0.65 A current from the wall charger or the battery pack as shown in Table~\ref{tab:rates_cap}. In other words, connecting Galaxy S2 with a higher output current charger does not enable faster charging. However, if a device and the charger both support fast charging, then charging the device with that charger would make the charging faster.  Charging a smartphone battery with higher than 1 A current is also called Fast charging. For example, our measurements show that Samsung Galaxy S4 draws 1.56 A current at 5 V from the charger during the CC phase of charging. In the case of a USB2.0 charger, the charging controller  draws a maximum 426 mA current.

Note that the current drawn by the charging controller ($I_{chg}$) may not be equal to the current pushed to the battery pack ($I_{bat}$). Along with the $I_{chg}$, we also measured $I_{bat}$ by placing the Yoshoo power monitor between the device and the battery. We measured $I_{bat}$ while a device is (1)  turned OFF, (2) idle and in airplane mode, and (3)  actively used. 
 The measurement results suggest that for the first case, $I_{bat} = I_{chg}$. For the latter cases, $I_{bat} = (I_{chg}-I_{sys})$. $I_{sys}$ depend on the power consumption characteristics of different hardware components being used. If the system is in airplane mode, then $I_{sys}$  vary within 10 mA, which is the standby current consumption of the device in airplane mode. If the device is actively used, then $I_{bat}$ decrease as the power drawn by different subcomponents, such as display, of the system increases. Consequently, the  battery will be charged slowly. If the system load is higher than the charger output current, the system drains the battery as well given  that battery has sufficient charge. Otherwise, the system shuts down. This happens when charging a device via USB2.0 charger. However, the system load does not change the current drawn from charger. 

\subsection{Charging New Batteries}
\label{sub:chargenew}
In this case, the smartphones were charged with their standard charger, cables, and new batteries. The initial battery capacity of Galaxy S2, S3, and S4 are 1650, 2100, and 2600 mAh respectively. These three batteries were manufactured in September 2014 and first used in this experiment. 

\begin{figure*}[t]
  \begin{center}
  \subfigure[GS4 Battery Voltage Curves with AC]{\label{fig:gs4_subvoltages}\includegraphics[width=0.32\linewidth,height = 0.3\linewidth]{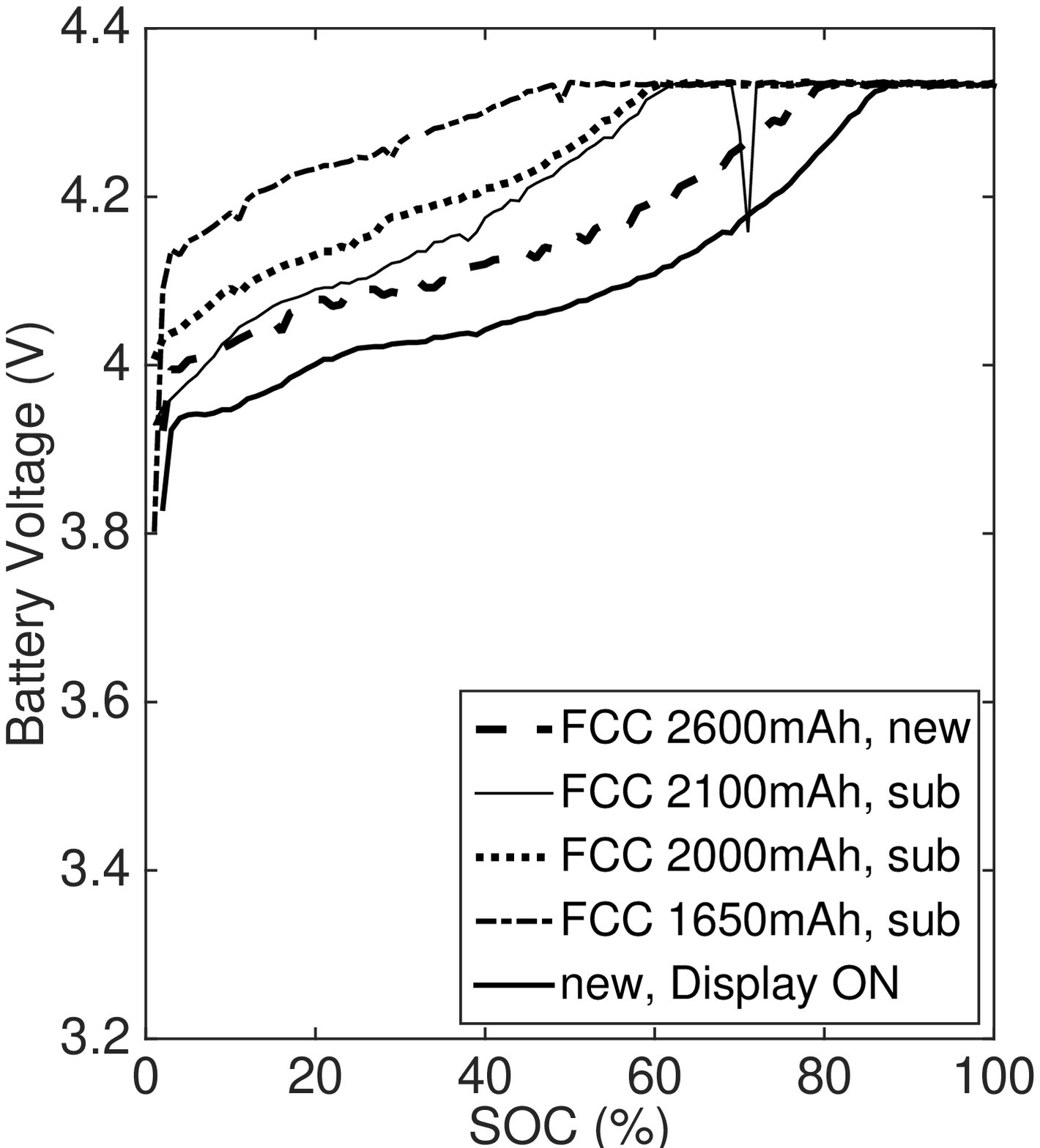}}
  \subfigure[GS3 Battery Voltage Curves with AC]{\label{fig:gs3_subvoltages}\includegraphics[width=0.32\linewidth,height = 0.3\linewidth]{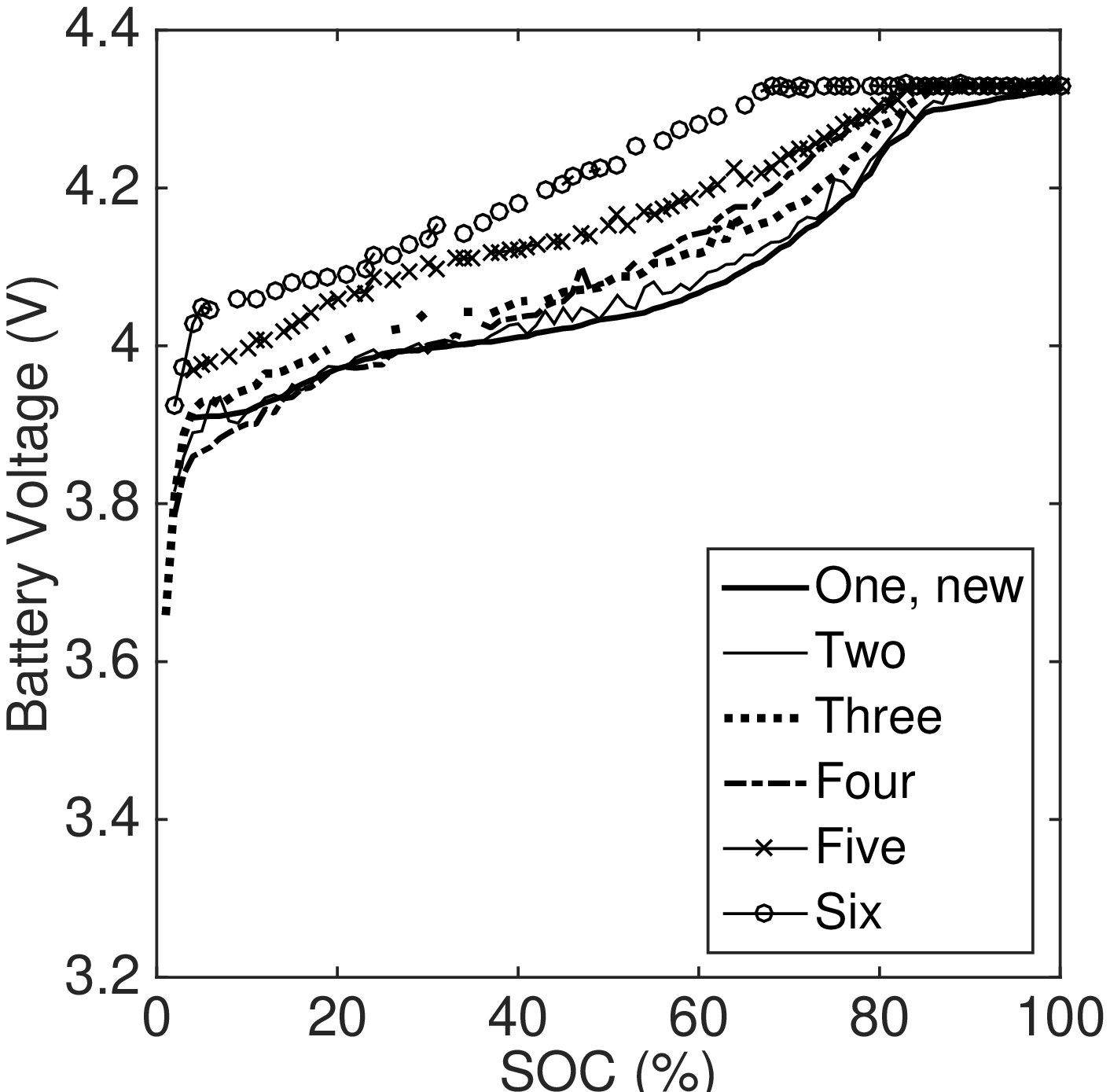}}
\subfigure[GS3 Battery voltage and charging current with AC]
{\label{fig:gs3_cvrates}\includegraphics[width=0.34\linewidth,height = 0.32\linewidth]{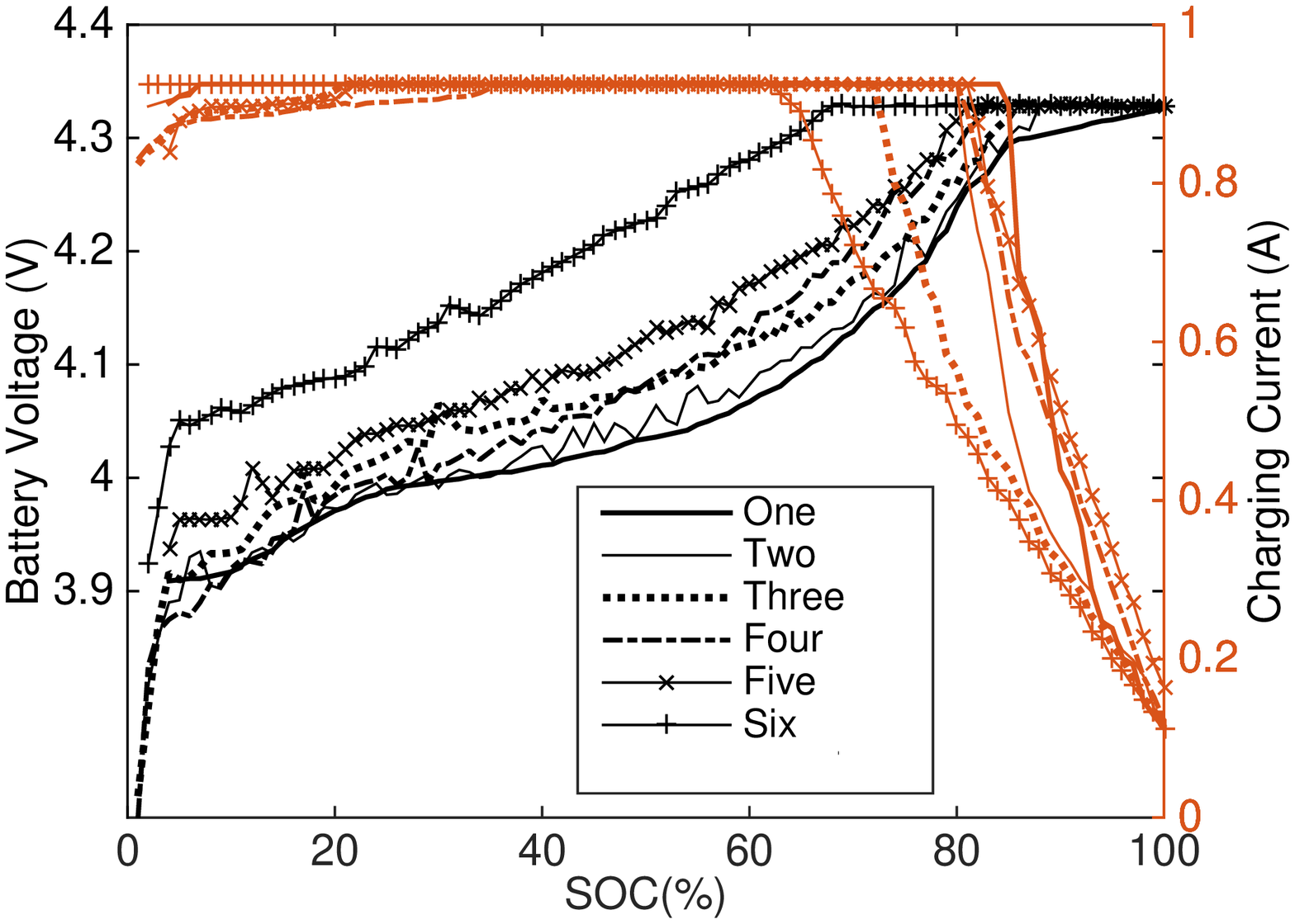}}

    \caption{{\bf Battery Voltage and charging current vs SOC while charging via AC.} {\sl Voltage curve of a lower FCC battery deviates from the curve of a new battery. The charging current plots (orange color) follow the legend of the corresponding voltage curves.}}

    \label{fig:gs4_charging}
 \end{center}
 \end{figure*}

Figure~\ref{fig:gs4_subvoltages} and ~\ref{fig:gs3_subvoltages} illustrate the relationship between battery voltage and the SOC while charging new batteries on Galaxy S3 and Galaxy S4. 
The battery voltage first increases sharply within battery level five and then the voltage increases almost linearly as the SOC increases over the remaining CC phase. This is because of feeding a constant current to the battery pack during the CC period. After that the battery voltage remains almost constant during the CV phase as the current is trickled. The SOC level that terminates the CC phase varies with the device and the corresponding SOC levels are 74, 85, and 76 for Galaxy S2, S3, and S4 respectively. 
We did not observe any such events of sudden drop in the battery level during these experiments that we did during the discharging experiments in \cite{mhotpower2015}. 


 Figure~\ref{fig:gs4_subvoltages} illustrates the voltage behavior when the new battery of Galaxy S4 is charged in airplane mode and when the display was ON with a fixed brightness level. We notice that the battery voltage of Galaxy S4 battery increases slowly compared to the case when the device is idle in airplane mode.

\subsection{Charging Lower Capacity Batteries}

To understand the behavior with non-standard batteries, we took out the new battery of Galaxy S4 and experimented with new substandard batteries having capacities of 2100, 2000, and 1650 mAh. These batteries were manufactured in September 2014 and first used in this experiment. 

We make several observations from Figure~\ref{fig:gs4_charging}.
First, the figure clearly shows that voltage varies for the same SOC with the new and old batteries. For example, when the SOC is 60\%, the observed voltages are 4.2 and 4.33 V for the new and the 2100 mAh battery respectively. As the battery capacity of Galaxy S4 decreases, the voltage of the battery increases for the same SOC.  Therefore, there is a unique charging voltage curve for each battery with different capacity (see Figure~\ref{fig:gs4_subvoltages}). 
Second, Figure~\ref{fig:gs4_subvoltages} further shows that the voltage of an old reaches its maximum value earlier compared to the charging of the battery with initial capacity.  Finally, there were small incremental jumps in the battery level. For instance, when the FCC was 2100 or 2000 mAh, the level increased from 96 to 100\% in a very short time. In the case of 1650 mAh battery, the device never completed charging. When the battery level reached 95\%, the device started discharging slowly even though the BatteryManager was broadcasting charging updates with decreasing SOCs.

\subsection{Charging Old Batteries}
\label{subsec:gs3_charging}


The measurements presented in the earlier section captures the behavior of battery voltage with batteries having less than the normal amount of capacity. 
We continue the investigation using old batteries in order to make sure that our observations are not an artifact of using non-standard batteries. We collected five old Galaxy S3, five Galaxy S4, and one Galaxy S2 batteries from our colleagues. These batteries were from one to three years old.

Figure~\ref{fig:gs3_subvoltages} compares the voltage curves of old batteries with the new battery. We see that battery voltage per SOC of the older  batteries is higher than that of the new batteries. The old batteries reach the maximum voltage at lower SOCs compared to a new battery. In other words, the older batteries have differing magnitudes less capacity, and the behavior in terms of battery voltage  is consistent with the decrease in FCC. We also observed similar patterns with USB charging. 





\subsection{CC-CV and Charging Current as the Battery Ages}
\label{subsec:yoshorate}

In this section, we verify the charging algorithms used by the devices with reduced battery capacity. During the charging measurements, we also instrumented the smartphone with Yoshoo. Since, the device does not have the functionality to export the measurements, we recorded the charging current measurements manually after every 60s. Afterwards, we associated the current with the voltage from the BatteryManager by associating SOC update times with the Yoshoo charging time in seconds. 

    

Figure \ref{fig:gs3_cvrates} shows the measurement results for Galaxy S3 batteries while the device was in airplane mode. We note that in some cases the charging current increases from an initial 800 mA to a stable 925 mA at the beginning of the CC phase. This charging pattern during the CC phase was also present with the Galaxy S2 and S4. We also notice that charging current begins to decrease when the voltage reaches its maximum value. The only exception is the battery $B3$ for which the current begins to decrease at SOC 73\% but the voltage reaches its maximum value when SOC is 80\%. This behavior persisted across all the measurements with battery $B3$. In the case of other Galaxy devices, the voltage behavior was in accordance with the charging current. Therefore, the behavior of $B3$ is battery specific. The charging rates are almost constant for the devices until they are charged to maximum 74\%, 85\% and 76\% respectively.

\subsection{Summary}
The presented voltage curves are the averages of per SOC measurements from four charging events. The voltage measurements varied by $\pm$0.05 V. The measurement results presented in this section lead us to the following conclusions.

\vspace{1mm}
\noindent\textbf{First}, \textit{The charging controller hosted in the device and the system load dictate the maximum charging current received by the battery.} Therefore, charging a device with a charger with higher output current does not enable fast charging unless both the charging controller and the charger support quick or turbo charging.  Charging the battery while a device is turned OFF or in complete idle state ensures the maximum current from the charging controller for the battery. If the device is actively used, the battery does not receive a constant maximum charging current due to the varied system load.

\begin{figure}[!h]
  \begin{center}
\includegraphics[width=0.8\linewidth,height = 0.4\linewidth]{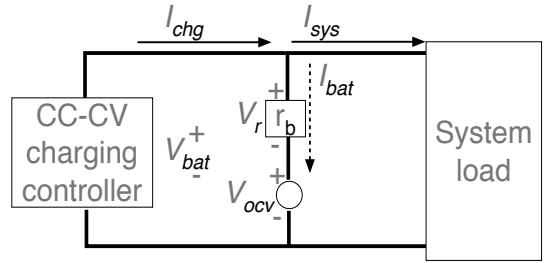}
    
    \caption{{\bf Equivalent circuit diagram while charging a battery.}}
    \label{fig:internal_res}
 \end{center}
 \end{figure}

\vspace{1mm}
\noindent\textbf{Second}, \textit{A battery with reduced capacity exhibits different behavior compared to a new battery without system load.} The voltage for a specific SOC of a battery with reduced capacity is higher than that of a new battery until the CC phase ends, and the larger the capacity loss, the smaller the SOC level at which the CC phase of charging is terminated and the voltage reaches its maximum value. The charging current pattern follows the battery voltage. However, the reason behind such voltage behavior of the aged batteries is attributed to their internal resistance. The internal resistance of a battery increases as the battery ages and temperature decreases. Figure~\ref{fig:internal_res} represents the equivalent circuit for charging a  battery. While a battery is charged, $V_{bat}$ increases from open circuit voltage ($V_{ocv}$) as much as a voltage drop across the internal resistance $r_{b}$, i.e., $V_{r}$. Since $I_{chg}$ is constant, $V_{r}$ becomes higher with lower temperatures or with aged batteries due to an increase of the battery's internal resistance. The effect of temperature on battery voltage and capacity is discussed in Section~\ref{subsec:battemp}.

\vspace{1mm}
\noindent\textbf{Third},\textit{ Mobile devices provide more reliable SOC estimates while charging compared to the discharging scenarios presented in ~\cite{mhotpower2015}}. We observed small incremental changes in the SOC at the end of CV phase while charging and the reason for better performance is that there is always some incoming energy from the charger.

\section{Full Charge Capacity Modeling and  Estimation}
\label{sec:fcc}
The state-of-art approach used by modern smartphones to estimate FCC is the number of charging cycles~\cite{Badam:2015}. Modern fuel gauges use Coulomb counter to measure the FCC of the battery. On the contrary, we propose a new FCC estimation technique based on the smartphone charging principle. In this section, we first present a charging rate-based FCC estimation technique and validate the method. We next derive a technique to estimate  the relative charging rate to estimate the FCC. First we validate the method for the devices with voltage-based fuel gauge and next we show that similar technique also works for Coulomb-counter based devices, such as Nexus 6. In this section, we consider charging device-specific standard batteries while the devices are in airplane mode.

\subsection{Full Charge Capacity Modeling and Validation}
As C-rate is the ratio between FCC and the charging current, it is possible to compute the present FCC of a battery from the C-rate. The  equation to compute C-rate from the battery initial capacity and charging current is the following. 

\begin{align}
\label{eq:crate}
C_{new} = \frac{I_{bat}}{FCC_{new}}
\end{align}

In Figure~\ref{fig:gs3_cvrates}, we have shown that the length of the CC phase reduces as the capacity of the battery reduces. However, the charging current (mA) drawn from the charger during the CC phase of charging does not change as the FCC of the battery decreases. Therefore, the $FCC_{now}$ can be defined as  

\begin{align}
\label{eq:crate2}
C_{now} = \frac{I_{bat}}{FCC_{now}}
\end{align}

\begin{align}
\label{eq:crate3}
FCC_{now}\times C_{now} = FCC_{new}\times C_{new}
\end{align}
\begin{align}
\label{eq:crate4}
\frac{FCC_{now}}{FCC_{new}} = \frac{C_{new}}{C_{now}}
\end{align}

\begin{table}[t]
\begin{center}
  \caption{Comparison of AC and USB2.0 charging C-rates, and FCC measurements and model estimates of FCC of batteries with different capacities.}

    \begin{tabular}{|p{11mm}|p{12mm}|p{10mm}|p{12mm}|p{12mm}|}
      \hline      
       Smartphone\break Model&$C_{new}$\break(ac, usb2.0)&$FCC_{now}$\break (loss \%)& $C_{now}$\break(ac, usb2.0) & $FCC_{now}$\break(ac, usb2.0)\break Model\\\hline
       GS4~(B1) & 0.6, 0.164& 2464~(5)& 0.63, 0.172 & 2476, 2479\\\hline 
       GS4~(B2) & 0.6, 0.164& 1042(60)&1.47, 0.408  &1061, 1056\\\hline
       GS4~(B3) & 0.6, 0.164& 1573(40)& 0.99, 0.270&1576, 1596\\\hline
       GS4~(B4) & 0.6, 0.164& 2048(21)& 0.76, 0.207 & 2053, 2090\\\hline
       GS4~(B5) & 0.6, 0.164& 1751(33)& 0.89, 0.243&1753, 1755\\\hline
       GS4~(B6) & 0.6, 0.164& 1763(32)& 0.88, 0.241 & 1772, 1769\\\hline

       GS2~(B1) & 0.39, 0.259& 1748(-6)&0.37, 0.243& 1739, 1758\\\hline 
       GS2~(B2) & 0.39, 0.259& 1665(-1)& 0.39, 0.255& 1650, 1675\\\hline 
       GS2~(B3)& 0.39, 0.259& ~613(40)& 1.05, 0.693& ~~613, 617\\\hline
       GS3~(B1)& 0.44, 0.202& 1992~(5)& 0.46, 0.214&2008, 1982\\\hline 
       GS3~(B2)& 0.44, 0.202& 1750(17)&0.53, 0.243& 1749, 1745\\\hline 
       GS3~(B3) & 0.44, 0.202& 1571(25)&0.59, 0.271&1566, 1565\\\hline 
       GS3~(B4) & 0.44, 0.202& 1703(19)&0.54, 0.250& 1711, 1697\\\hline
       GS3~(B5) & 0.44, 0.202& 1491(29)&0.62, 0.286&1490, 1483\\\hline
       GS3~(B6)& 0.44, 0.202& 1083(48)&0.85, 0.392&1087, 1082\\\hline
  \end{tabular}
         \label{tab:rates_info}
    \end{center}
 
\end{table}

And consequently, the present capacity of the battery can be computed with \eqref{eq:crate3}. This reveals that present battery capacity of the battery is the ratio of the charging C-rates, i.e.,  the ratio between the charging rate with present unknown capacity with the charging rate of the new battery. This is shown in equation \eqref{eq:crate4}. In the above equations, the C-rate of a new battery, $C_{new}$, can be derived from the battery capacity and charging current information.


In order to find the C-rate of the new and old batteries, first we need to measure the capacity of the batteries. The batteries usually come with labeled capacity values. The  charging C-rates of the new batteries of the Galaxy S2, S3, and S4 are 0.39(645/1650), 0.44(925/2100), and 0.6(1560/2600) C respectively when charged via the AC charger. Table~\ref{tab:rates_info} also shows the charging C-rates with USB2.0 charger. 

We measured the FCC of the batteries by discharging them at 0.8 A rate. The results presented in Table~\ref{tab:rates_info} reveal that the batteries have 5-60\% less capacity than the labeled value. There are a couple of other interesting observations from table~\ref{tab:rates_info}.  First, the new batteries of Galaxy S3 and S4 have less capacity than the label indicates. We can think of two reasons for this; either a battery may actually come with less capacity, or the battery does not allow itself to be discharged completely and therefore there is always some small amount of charge remains in the battery. Second, one of the new batteries of Galaxy S2 has 100 mAh more capacity than the labeled value. 

\begin{figure}[t]
  \begin{center}
  \subfigure[Galaxy S3]{\label{fig:gs3_subrate}\includegraphics[width=0.48\linewidth,height = 0.5\linewidth]{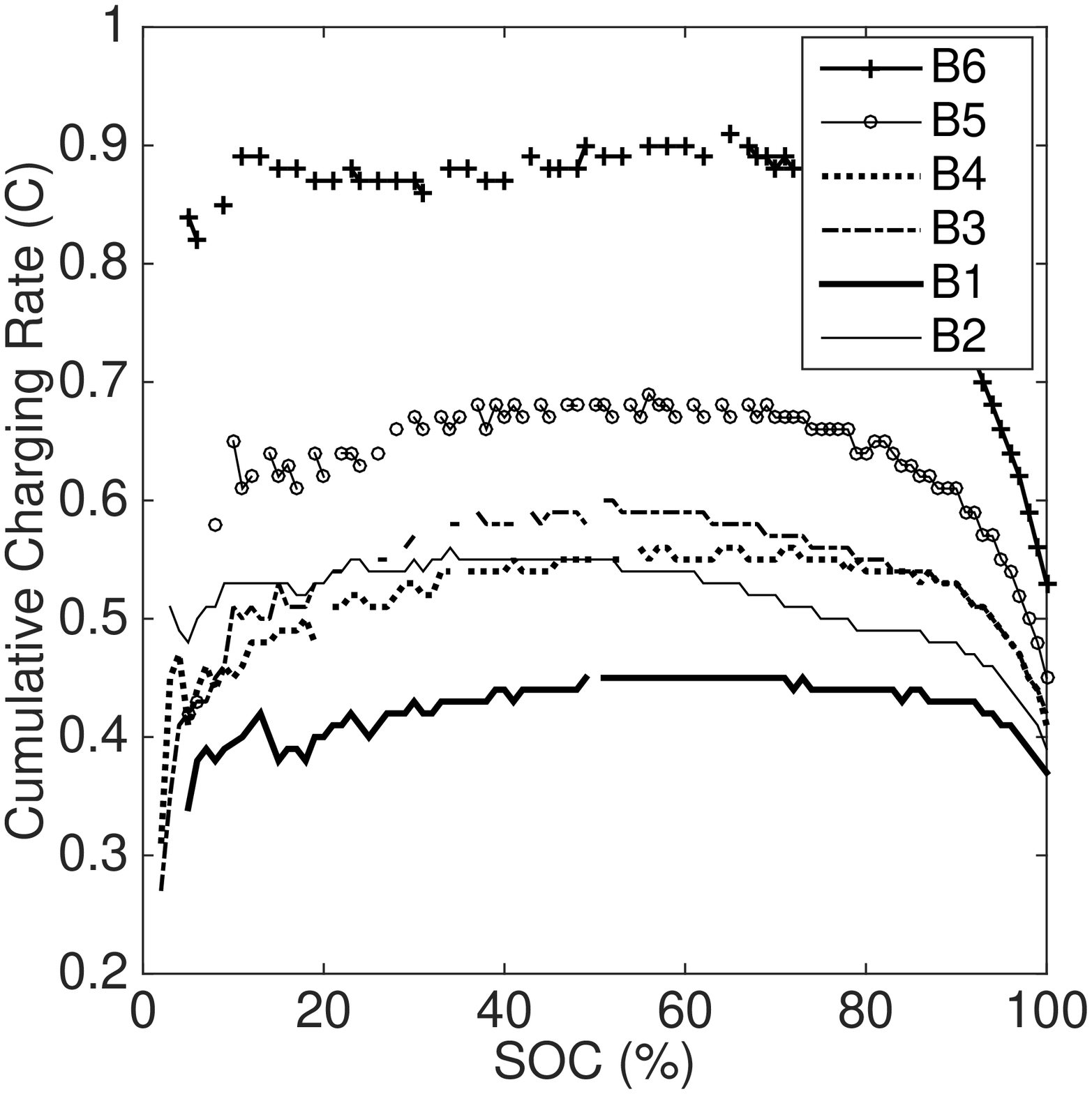}}
  \subfigure[Galaxy S4]{\label{fig:gs3_subrates}\includegraphics[width=0.48\linewidth,height = 0.5\linewidth]{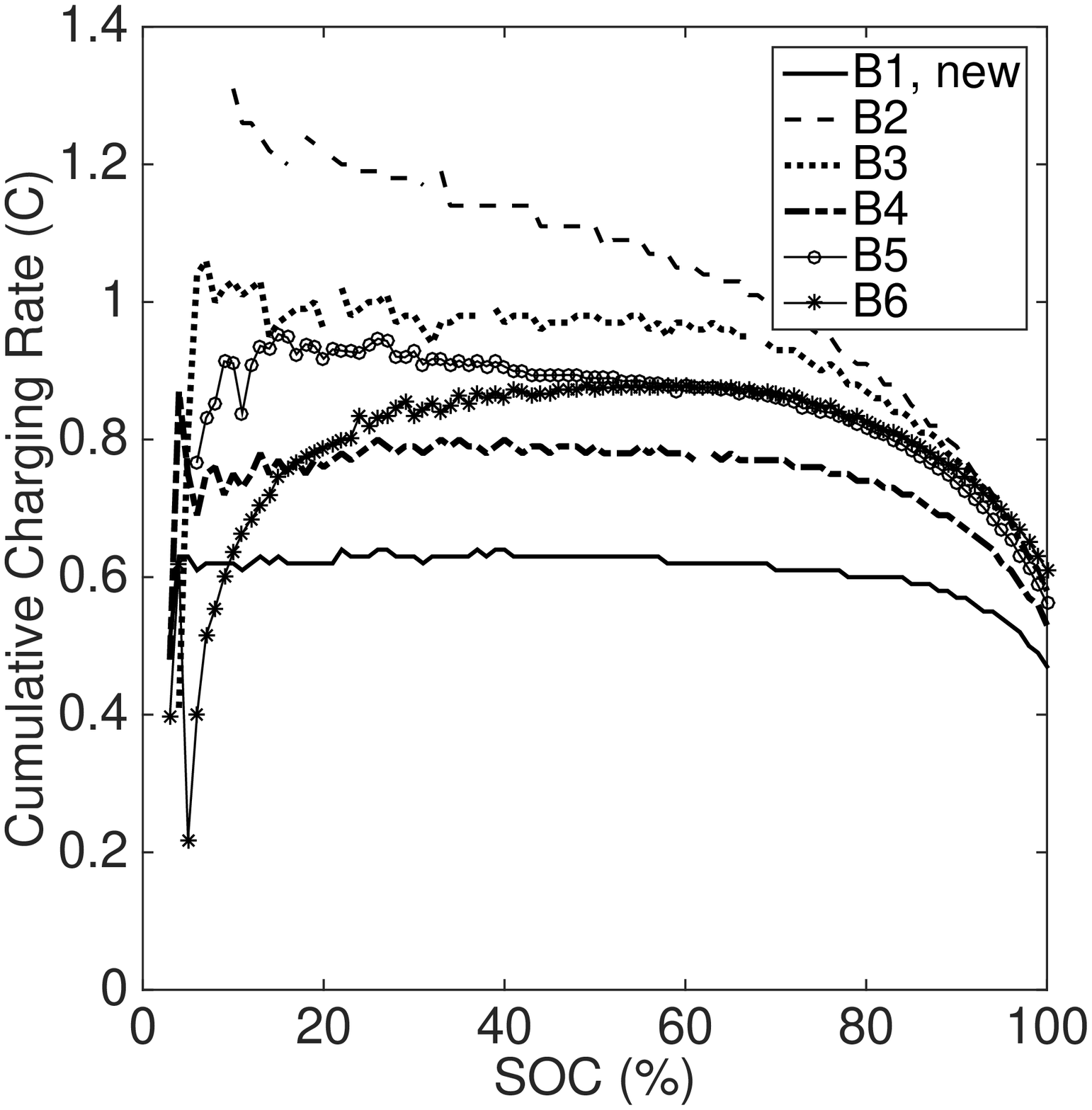}}\\
    \caption{{\bf The cumulative AC charging rate curves of the Galaxy devices with new and long used batteries. }{\sl The charging rate increases as the capacity decreases.}}
    \label{fig:all_charging}
 \end{center}
 \end{figure}

\begin{figure*}[t]
  \begin{center}
  \subfigure[Galaxy S3 Batteries]{\label{fig:gs3fcc_compare}\includegraphics[width=.32\linewidth,height = 0.3\linewidth]{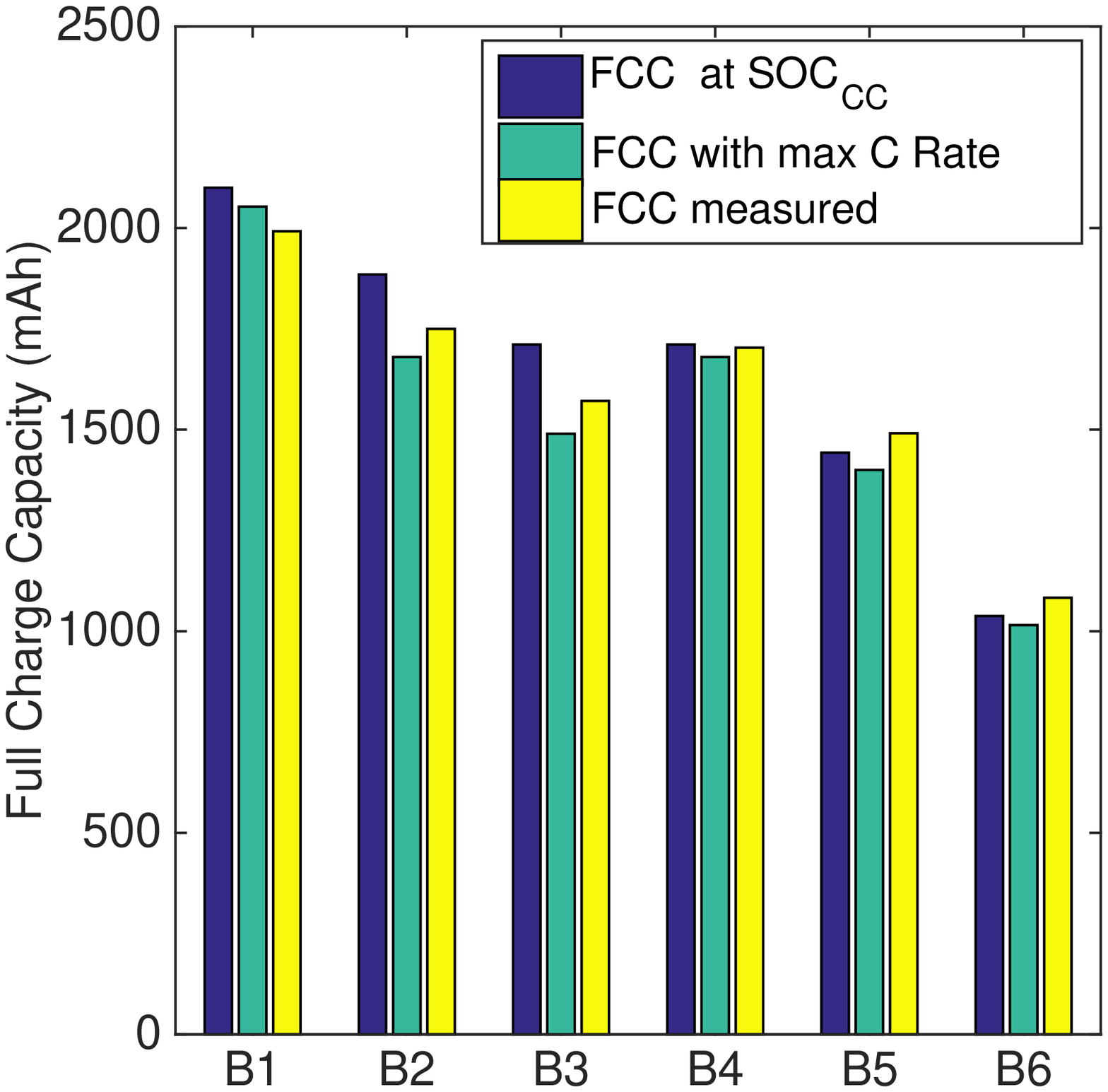}}
\subfigure[Galaxy S4 Batteries]{\label{fig:gs4fcc_compare}\includegraphics[width=0.32\linewidth,height = 0.3\linewidth]{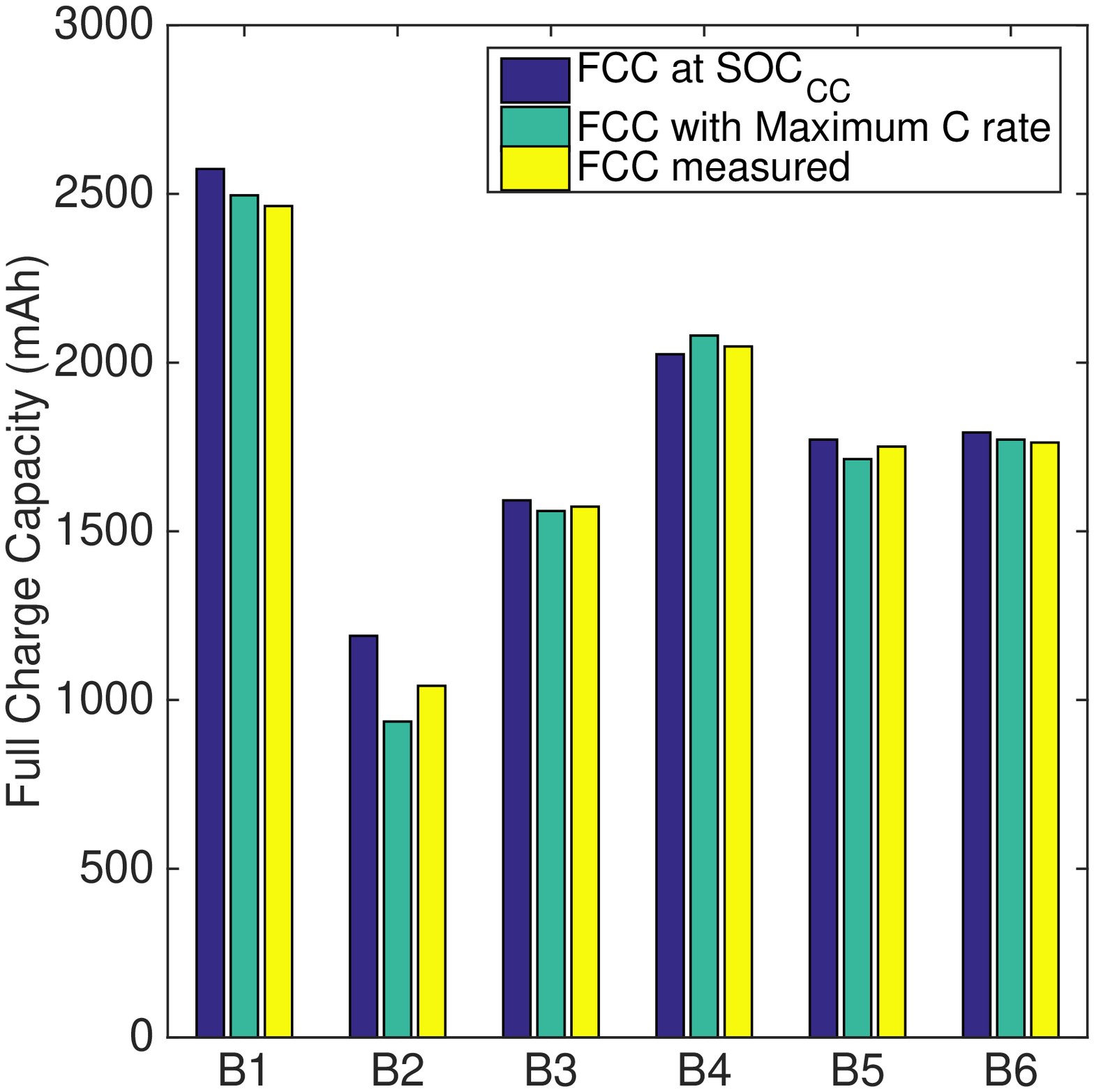}}
\subfigure[Galaxy S3 Batteries]{\label{fig:gs3perecent_compare}\includegraphics[width=0.32\linewidth,height = 0.3\linewidth]{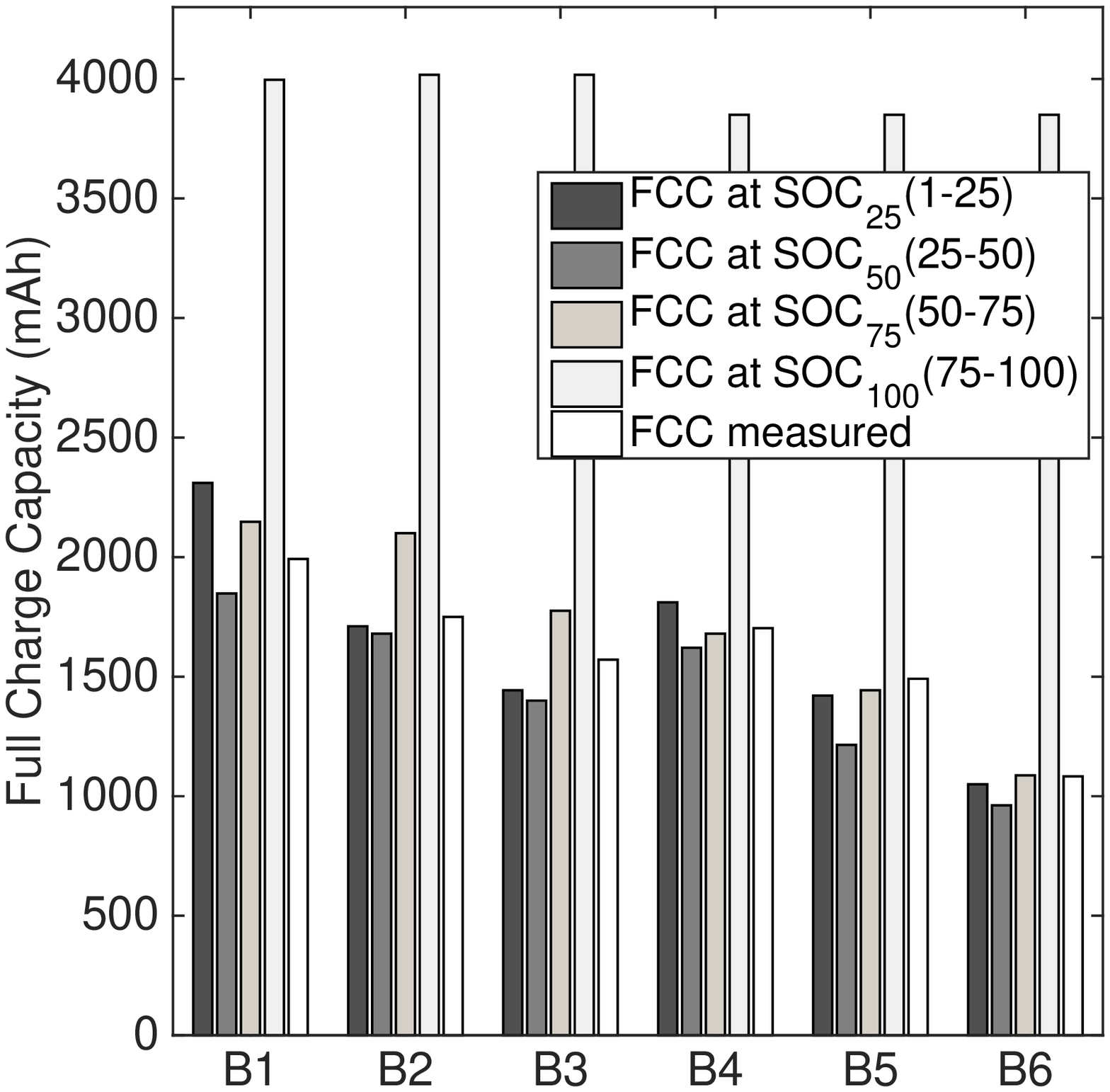}}

    \caption{{\bf Comparing the FCC estimation of Galaxy devices' batteries with the capacity measurements. }{\sl The smallest CC phases for the S3 ($B6$), and S4 ($B2$) batteries are 68 and 10 SOC respectively. The estimates with different rates are close to the measurement results, except with the rates during the CV phase.}}
    \label{fig:fcc_compare}
 \end{center}
 \end{figure*}

We next estimate the charging C-rates of the batteries according to eq.\eqref{eq:crate2} for both AC and USB2.0 charging. The results are presented in Table~\ref{tab:rates_info} and we notice that the charging C-rates increase as their capacity decrease. Finally, we estimate capacity using eq.~\eqref{eq:crate4} and compare with the measurement results in the table. The table shows that the model estimates the FCC quite reasonably for both the AC and USB2.0 charging.


\subsection{C-rate from SOC Updates}
\label{sub:battcrate}
In the earlier section, we measured the FCC of the batteries, and from the measurements we estimated the $C_{now}$, and finally validated the model estimated by comparing with the measurement results. Given the initial battery capacity and the charging current, $FCC_{now}$ and $C_{now}$ are two unknown variables, which depend on each other. Therefore, we need to find the $C_{now}$ in order to compute the $FCC_{now}$. In this section, we devise a method to estimate $C_{now}$. 
The definitions of the charging algorithms state that CV phase starts when the battery voltage reaches its predefined maximum value and during CC period the charging current remains constant.  Therefore, we can estimate the rates from the time stamp of the same SOC updates for the charging measurements presented in Section~\ref{sec:three}. The equation is the following,

\begin{align}
\label{eq:one}
C_{SOC_{i\rightarrow n}}=\frac{36\times (SOC_{i+n}-SOC_{i})}{t_{SOC_{i+n}}-t_{SOC_{i}}}~,
\end{align}

\noindent where 36 is the time to charge one percent at 1 C-rate. Using equation \eqref{eq:one}, we can estimate the cumulative charging rate over a SOC interval, such as the C-rate to charge from 2 to 50\%. We compute the cumulative charging rates of the devices and plot in Figure~\ref{fig:all_charging}. If we take the rates of the devices at the point where the CC phase ends, we find that AC charging rates of the new batteries of Galaxy S2, S3, and S4 are 0.38, 0.44, and 0.59 C respectively which are very close to the measured rates presented in Table~\ref{tab:rates_info}.  The figure also highlights that although the charging rate from the wall charger is almost constant during the CC phase irrespective of the battery capacity, the C-rates of the older batteries are higher than the new batteries as derived earlier through measurements. However, when the display was switched ON, the Galaxy S4's charging rate was 0.5 C, which exemplifies the bias due to the device utilization  while charging.

\subsection{FCC estimate from  Charging SOC C-Rate}
Now from the C-rate curves presented earlier we need to select a rate within the CC phase SOC boundary, which would reflect the $FCC_{now}$ of the battery. However, it is a challenge to select the range of SOC values over which the C-rate should be calculated given the rate curves presented in Figure~\ref{fig:all_charging}. We notice that the estimated C-rates are not constant during the CC phase. For instance, although the CV phase of battery $B2$ starts when the SOC is 80\%, we notice that the battery is charged with the maximum C-rate till 40\% and after that C-rate gradually decreases. 

In the case of battery $B4$, the max C-rate is observed when the SOC is 80\%. One possible explanation is that although the batteries are charged at a constant rate from the wall charger, charging an individual SOC level may take different amount of time.  We experimented  the batteries over four charging events and found that this is individual battery characteristic. Because of this battery specific behavior, we explore and validate two different options; 1) we select the range of SOC values that cover the whole CC phase or 2) we select the SOC range that yields the highest C-rate (we call this max C-rate). The lengths of the CC phase are derived from the voltage curve of the battery.

We next plug in the C-rate values in \eqref{eq:crate3}. Multiplying the C-rates ratio  with $F_{new}$ in mAh would give $F_{now}$ in mAh. We estimate and compare $FCC_{now}$ with two different C-rate options mentioned earlier. Figure~\ref{fig:gs3fcc_compare} and \ref{fig:gs4fcc_compare} compare the FCC measurement results with the model estimates of Galaxy devices.  We notice that both of the rate selection approaches estimate $FCC_{now}$ with less than 10\% error. The C-rate at $SOC_{CC}$ yields an FCC estimate which is closer to the measurements more often than the other approach. However, in most cases the model estimates are higher than the measurement results. The obvious reason is the tiny system load while charging the batteries. The SOC C-rates for these batteries with USB2.0 charging gives similar estimates. For example, the USB2.0 charging C-rates at $SOC_{CC}$ for $B3$ batteries of Galaxy S3 and S4 are 0.265 and 0.266 C  respectively and the corresponding FCC estimates are 1609 mAh and 1594 mAh.

Although the Android BatteryManager broadcasts SOC update events regularly, there may be only few updates available in practice. The underlying reason can be device-specific behavior in reporting SOC updates or that the device is in doze mode and unable to broadcast these events. In addition, a user may charge the device when the device is off and switch it on only after the device is charged to a reasonable capacity. Therefore, our method has to work also when it has only partial SOC updates available. Figure~\ref{fig:gs3perecent_compare} illustrates the FCC estimates at the boundaries of four different SOC intervals and compares with the measurement results. We notice that within 75\% SOC boundaries the FCC estimates are close to the measurement results. Beyond that SOC level, the error increases significantly due to the trickling charging current during the CV phase.


\begin{figure}[t]
  \begin{center}
  \subfigure[Galaxy S3 (B1)]{\label{fig:gs3voltage_temp}\includegraphics[width=0.48\linewidth,height = 0.5\linewidth]{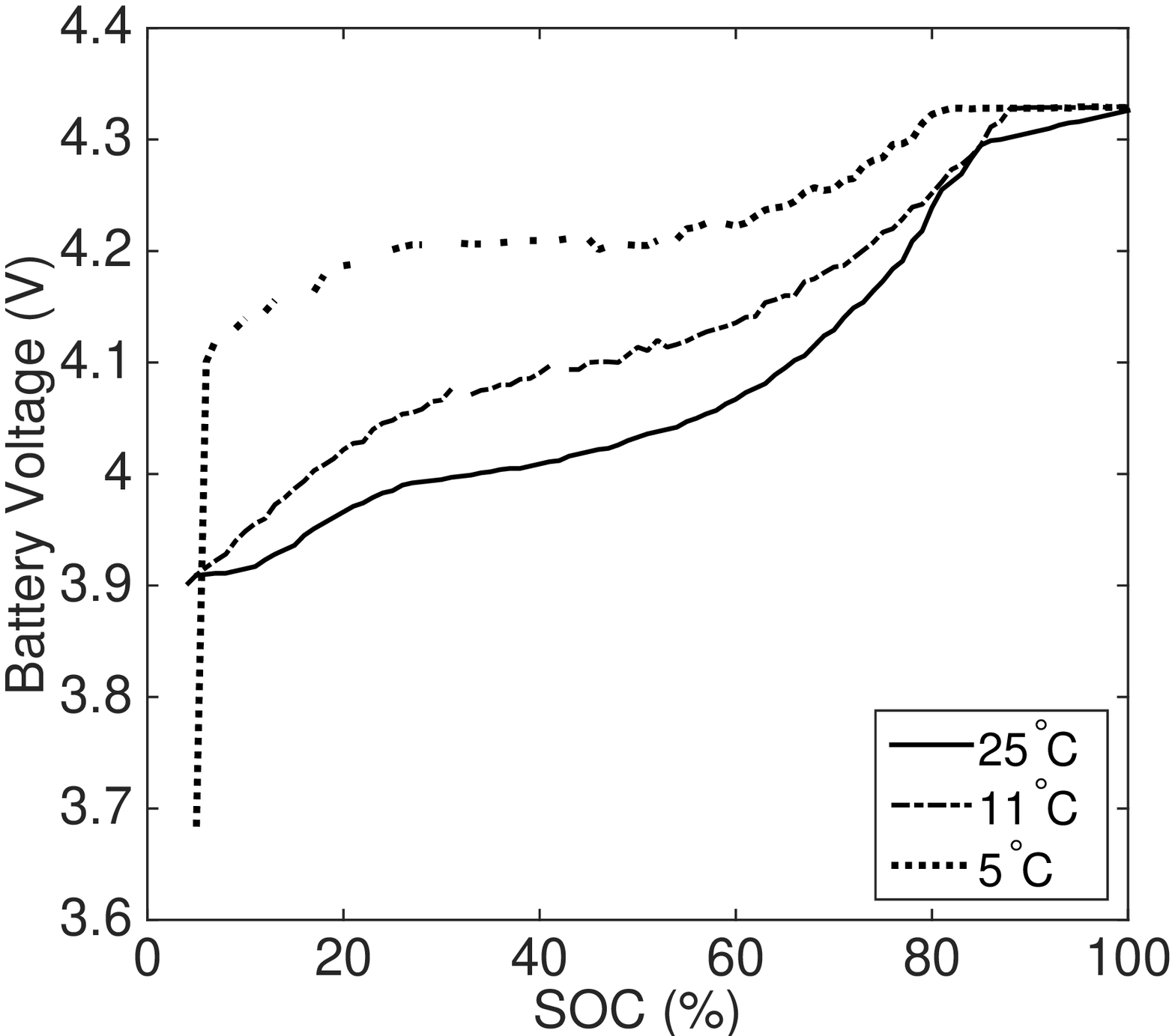}}
  \subfigure[Galaxy S3]{\label{fig:gs3temp_fcc}\includegraphics[width=0.48\linewidth,height = 0.5\linewidth]{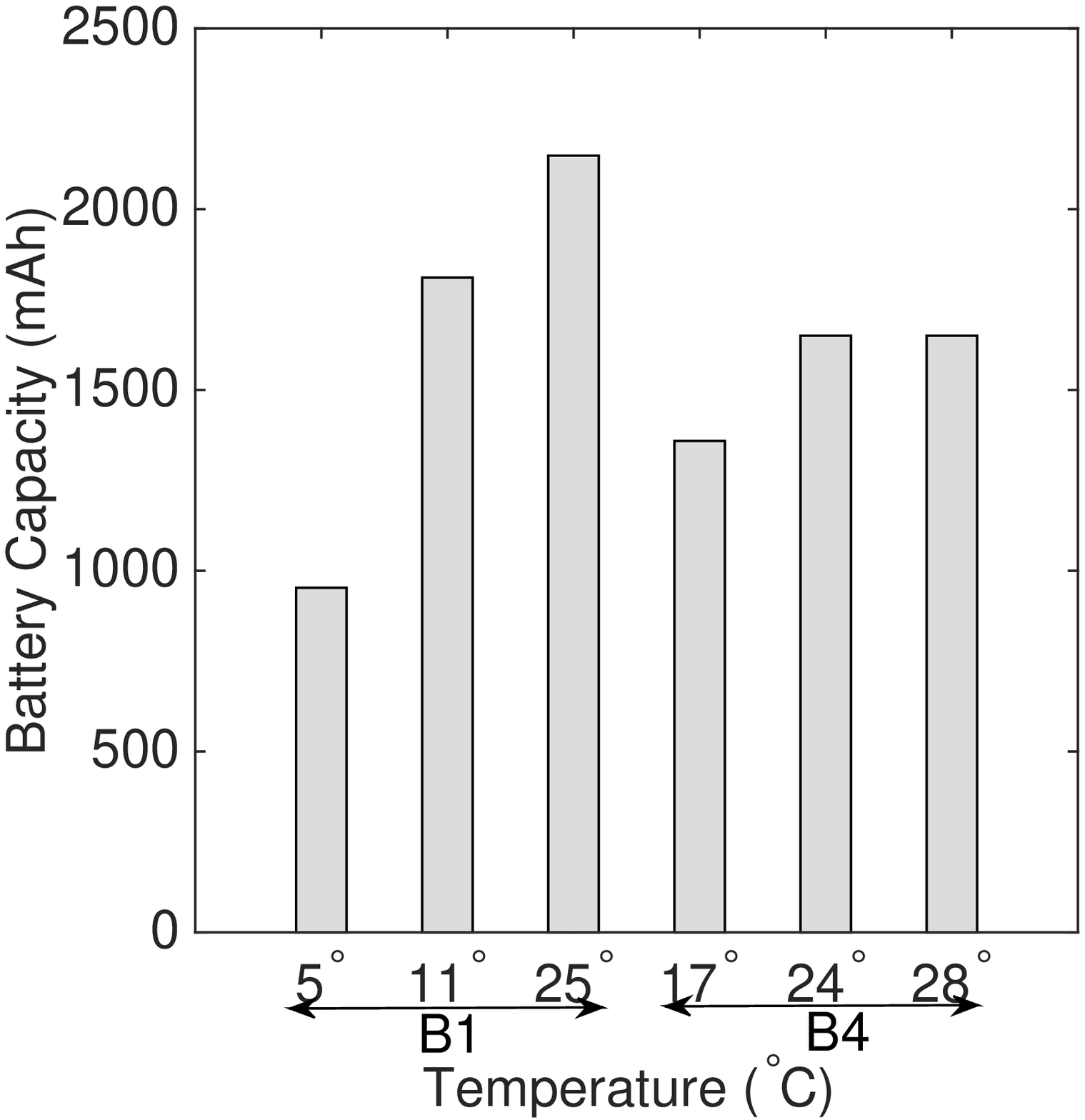}}\\
    \caption{{\bf The effect of temperature on Battery Voltage and Capacity while charging via AC. }}
    \label{fig:temp_charging}
 \end{center}
 \end{figure}

\subsection{FCC at Low Temperature}
\label{subsec:battemp}
We further conducted charging experiments with varying temperatures. Figure~\ref{fig:gs3voltage_temp} demonstrates voltage behavior of the new Galaxy S3 battery with different temperatures while charging. The voltage behave in a similar fashion to those of the old batteries. We compute the FCC from the SOC C-rates for the charging scenarios at different temperatures. Figure~\ref{fig:gs3temp_fcc} shows that the FCC of $B1$ and $B4$ decrease when they are charged at lower temperature than the room temperature. For example, $B1$ was charged to 1811 and 953 mAh at 11$^{\circ}$ and 5$^{\circ}$C. The battery $B4$ was charged to 1359 mAh at 17$^{\circ}$C. We further investigated whether the effect of such charging has short or long term effect on FCC. We discharged the batteries at room temperature, after we had charged them at lower temperature. Afterwards, we relaxed them for two hours, and then charged again at room temperature. We found that all the  batteries retained their earlier capacity state.  Such results suggest that temporary low temperature charging has short term effect on the battery capacity.

\subsection{FCC of a Coulomb Counting Device}
In the earlier sections, we measured and derived the FCC for devices with voltage-based fuel gauges. In this section, we consider Coulomb counter-based a new Nexus 6 device. The FCC of this device is 3010 mAh. The charging rates of the device with USB2.0, standard 5 V AC, and turbo 9 V AC charging are 0.144(440/3050), 0.46(1400/3050), and 0.78(2400/3050) C respectively. We next measured the battery voltage from SOC updates, and estimated the charging C-rate for three charging configurations. The battery voltage behavior is similar to those of the Galaxy devices, however the battery voltage reaches well above the maximum voltage configuration 4.35 V for both the standard and turbo charging. The lengths of the CC phases are 70, 80, and 95 SOCs. The charging terminates when the voltage reaches to 4.35 V. The corresponding charging rates are 0.143, 0.45, and 0.77 C respectively at the end of the CC phases. These estimated C rates provide battery capacities  3009, 3050, and 3086 mAh respectively which are very close to the measured capacity.

\begin{figure}[t]
  \begin{center}
\includegraphics[width=1.0\linewidth,height = 0.5\linewidth]{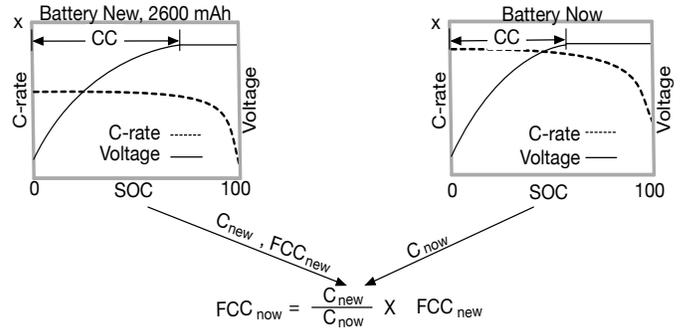}
    
    \caption{{\bf The comparison of Voltage and rate curves of a battery  towards finding the FCC of the battery online.}}
    \label{fig:cratemethod}
 \end{center}
 \end{figure}

\begin{table}[t]
\begin{center}
       \caption{Galaxy S2, S3, S4 users and their FCC losses in a crowdsourced battery analytic data set. Most of the devices had maximum 25\% FCC loss.}
           
  {\footnotesize
    \begin{tabular}{|p{8mm}|p{9mm}|p{9mm}|p{9mm}|p{9mm}|}
      \hline      
        {Model}& { Users}& {25}\%&{50}\%&{75}\% \\\hline
GS2&421&166&17&	2\\\hline
GS3&284&149&21&	4\\\hline
GS4&318&136&24&	6\\\hline
  \end{tabular}}
          \label{fig:crowd_results}
    \end{center}
 
\end{table}

\subsection{Summary}
We have demonstrated that the relative charging rate within the CC phase of a battery increases as the FCC decreases and based on this C-rate we have also proposed an FCC estimation model. It can be enforced with partial  charging battery updates from an uninterrupted charging session and accuracy of the model is above 90\%. The FCC estimation technique can be summarized  according to Figure~\ref{fig:cratemethod} given that the smartphone is idle while charging. The method works with different SOC estimation techniques. Among the rate selection methods described earlier, we select the C-rate at $SOC_{CC}$ and apply this with the users in the crowdsourced Carat data set and compare with the C-rate of the Galaxy devices with new batteries. We have also shown that the model works for different temperature settings and with both Coulomb-counting  and voltage-based fuel gauges. Since our method depends on the SOC updates and voltage measurements from the battery, the performance also depends on the accuracy of the SOC estimation techniques and we have further shown that performance of these techniques is more accurate while charging. Table~\ref{fig:crowd_results} describes the capacity loss of users devices. We see that 43-50\% of the devices of these models suffered from capacity loss and a significant number of them lost 25\% of the capacity. We describe the Carat data set more detail  and the crowdsourced FCC  estimation method in Section \ref{sec:five}.

\section{Crowd-sourced Battery Analytics Data Set}
\label{sec:five}
The measurement results presented in Section~\ref{sec:three} demonstrate that battery voltage can characterize the FCC of the battery while charging via AC. The charging rate from the BatteryManager updates enables to characterize and compute the FCC more reliably (see Section~\ref{sec:fcc}). However, smartphones  do not report the capacity of a  battery and the charging rate. Therefore, in this section we devise statistical methods to find the $C_{new}$ and $C_{now}$ of an unknown device given a large collection of the Android BatteryManager data of a particular smartphone model.  This also enables online collaborative debugging of the smartphone battery FCC. We identify the devices with reduced FCC from the Carat data set. Carat is a collaborative mobile energy debugging framework~\cite{Oliner2013}. It has user applications for Android and iOS devices to collect the battery, application, system, and networking information from mobile devices.

\subsection{Data Set $\&$ Pre-processing Charging Samples}In this section, we first describe the data set and some pre-processing steps on the data to be used in the later sections. The Carat application collects different information as samples whenever there is a change in the SOC or battery level. A sample contains a lot of other information along with the battery related information and can be defined as $S = (t,(a_{1}:v_{1}),(a_{2}:v_{3}),(a_{3}:v_{3})...(a_{n}:v_{n}))$, where $t$ is the epoch time stamp when the sample was captured, and  $(a_{i}:v_{i})$ are the attribute and value pairs. We took a subset of the data from January 2013 to February 2014 of 200 GB  consisting of 8 million samples from 42 K devices of 2 K unique models.

\begin{figure}[t]
  \begin{center}
  \includegraphics[width=1.0\linewidth,height = 0.4\linewidth]{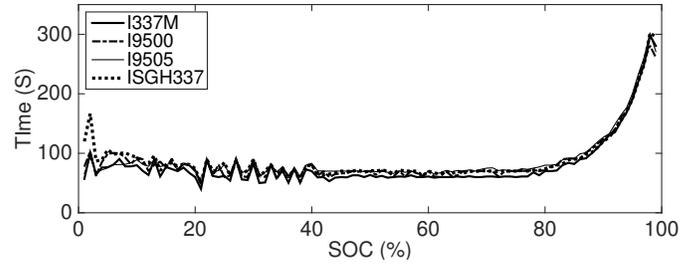}
  
    \caption{{\bf One percent charging time curves of various Galaxy S4 models. }{\sl One percent charging time is almost constant with respect to SOC within the CC phase (75\%).}}
    \label{fig:onePercurve}
 \end{center}
 \end{figure}

In order to construct the SOC vs battery voltage curves and the charging rate curves, as shown in the earlier sections, we rely on the AC charging samples from the data set. From the list of information in a sample, we only consider the time when the sample was taken, the SOC, battery voltage, battery health, the type of charger, and the CPU usage. Hence, the reduced sample looks as  $S = (t,~(SOC:i\%),~(voltage:V),~(temp:C)~(charger:ac/usb),~(health:good/dead/cold),~(cpu:x\%))$. First, we consider the samples with the \textit{charger} attribute of ``ac'' and health attribute of ``good'' value. As it is demonstrated earlier that charging a battery at lower temperature than the room temperature can affect the FCC,  we  consider  only those  samples with battery temperature reporting 21-40$^{\circ}$C, as the capacity variation within  this  range  is  very  small. There were about 3 million charging samples and about 22 K devices of 1200 models had more than 5 good charging samples. However, the samples did not have the display status, i.e., ON/OFF, of the devices during the sample collection.

We next sort the good AC charging samples of a user according to the time stamp in order to find the charging time between two consecutive samples. First we need group the samples that belong to same charging events. Ideally, a charging event begins when a charger is plugged in and ends when the charger is unplugged. However, the construction of the events in this way is difficult from the data set as a user may turn ON/OFF the phone while charging and turn ON when the battery is charged to a reasonable capacity. The charging algorithms terminate charging once the charging rate falls to 0.07~\cite{Thanh:2012}. Therefore a mobile device spends at most $\frac{36}{0.07}=514$ seconds to charge one percent.
We next add this derived attribute in the samples and finally we obtain the following kind of samples: $S = (t,~(SOC:i\%),~(voltage:V),~(\Delta t:S),~(cpu:x\%))$. 
All the pre-processing is done using Spark~\cite{Zaharia2010} with 7 machines each having 8 CPU cores and 30GB RAM.


\subsection{Methodology}
\label{sub:methodology}

Since we do not have battery vendor specific information, such as the battery capacity and the charging current drawn, our analysis relies on the observed characteristic presented in Section~\ref{sec:three} and~\ref{sec:fcc} and on the crowdsourced battery information.  Our method for inferring the FCC of the battery of a device compares the charging behavior of a user device against the behavior of a community of the same model devices.  Since the size of the battery for a specific model is unique, we can construct  model specific voltage and charging rate curves when we have sufficient number of users using  the same model and with a reasonable amount of samples. Our method relies on a number of steps. First, we find the model-specific voltage curve and the length of the CC phase. We next determine the model- and user-specific relative charging rates during the CC phase. These two rates are equivalent to the $C_{new}~\&~C_{now}$, respectively. Finally, we compare these two rates according to \eqref{eq:two} to determine the capacity loss.

Determining the model-specific voltage curves from the crowdsourced data is not trivial as it is difficult to say whether a device was completely idle during the sample collection. Since Carat is an energy debugging engine, the users can also be biased towards having devices with the lower FCCs. As we have demonstrated in Section~\ref{sec:three}, the effects of these two facts on battery voltage while charging are the opposite. Consequently, an individual sample can be biased. Therefore, we apply \textit{G-test}~\cite{kartusis} on battery level specific voltage distribution to determine the skewness with the confidence of $\alpha = 0.05$. Left skew implies samples biasness towards lower FCC devices. If the distribution is right skew and symmetric then the distribution is affected by the device usage. Therefore, if the distribution of an individual SOC is left skewed, we consider the median voltage, else we consider the 75th percentile.

From the corresponding voltage curve samples, we also construct the one percent charging time curves. Example charging time curves of four models is presented in Figure~\ref{fig:onePercurve}. The charging time curve of an individual device is constructed only from the their most recent available samples which characterizes the recent state of the battery. We later in Section~\ref{sub:fcc_loss}, use this charging time to compute the FCC. These charging voltage and time curves are processed and analyzed in Matlab. The steps are presented in Algorithm~\ref{alg:cevent} with Matlab notation. $mVCrv$ in Alg.~\ref{alg:cevent} represents the model specific voltage curve derived according to the earlier described steps.

\begin{figure}[t]
  \begin{center}
 \includegraphics[width=0.95\linewidth,height = 0.4\linewidth]{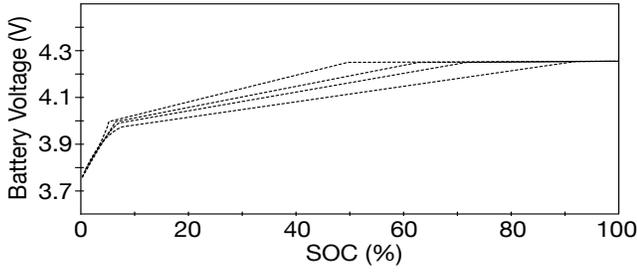}
    \caption{{\bf The voltage curve can be divided into three linear segments.} {\sl The slope of the second segment is lower than the first segment and the third segment is parallel to the X-axis.}}
    \label{fig:curve_seg}
 \end{center}
 \end{figure}


\subsection{Reference Voltage Curves}

The reference voltage curve should be close to the curve constructed with a new battery. Therefore, the number of samples per model should be sufficient and it is essential to have as many non-null SOC elements as possible in the curve. We select the device models, which had a minimum 250 samples. This gives us a wide coverage of different models. Among 370 models, we found that approximately 300  models had more than 90\% non-null SOC entries in their reference curves. The remaining device models had more than 60\% SOC entries per curve.


\begin{algorithm}[t]
\caption{Battery FCC Loss Detect and Estimate}\label{alg:cevent}
\begin{algorithmic}[1]
\Function{BatteryFCCEstimate}{}
\For{\textbf{each} \texttt{model} $\in$ \texttt{MODELS}}
    \State \texttt{mVCrv=mVoltageCurve(1:100)}
    \State \texttt{[mVCrv,mcc]=interpolateV(mVCrv)}
    \State \texttt{mVCrv=delOutliers(mVCrv(10:mcc))}
    \State \texttt{[mVCrv,mcc]=interpolateV(mVCrv)} 
    \State \texttt{mTCrv=mTimeCurveVoltage(1:mcc)}
    \State \texttt{mTCrv=delOutliers(mTCurve(1:mcc))}
    \State \texttt{mTCrv=Interpolate(mTCurve(1:mcc))}
    \State \texttt{mRateC=cumRate(mTCurve(1:mcc))}
    \State \texttt{mRate=mRateC(mcc)}
    \For {\textbf{each} \texttt{device} $\in$ \texttt{modelDEVICES}}
        \State \texttt{uVcrv=uVoltageCurve(1:100)}
        \State \texttt{[uVcrv,ucc]=interpolateV(uVCrv)}
        \State \texttt{uTCrv=uTimeCurveVoltage(1:ucc)}
        \State \texttt{uTCrv=Interpolate(uTCrv(1:ucc))}
        \State \texttt{uRateC=cumRate(uTCrv(1:ucc))}
        \State \texttt{uRate= uRateC(ucc)}
        \State \texttt{$\Delta r$ = uRate-mRate}

        \If{\texttt{($\Delta r>0$)} \&\& $(ucc > 0)$}
            \State \texttt {uCap = mRate/uRate}
            \State \texttt{uLoss=(1-uCap)}
        \EndIf
    \EndFor
\EndFor
\EndFunction
\end{algorithmic}
\end{algorithm}

\begin{figure*}[t]
  \begin{center}
  \includegraphics[width=1.0\linewidth,height = 0.2\linewidth]{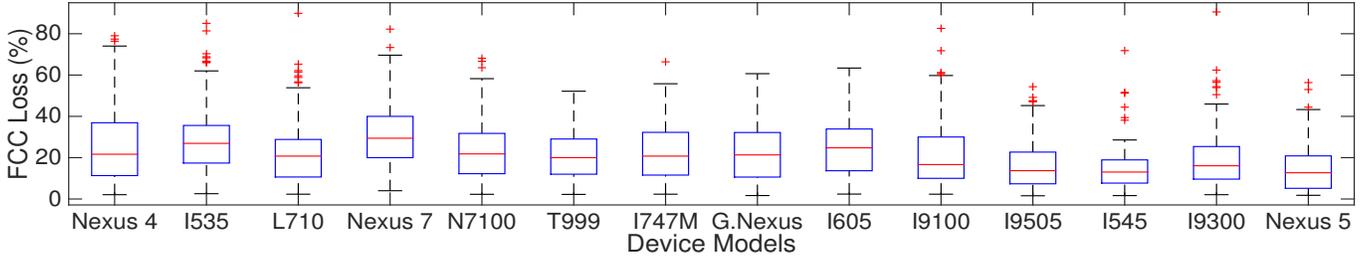}
    
    \caption{{\bf Capacity losses of devices of popular fourteen models. }{\sl Nexus 4, Nexus 7, Galaxy Nexus, and SCH-I605 devices had significant capacity loss.}}

    \label{fig:devloss}
 \end{center}
 \end{figure*}

However, the 250 samples do not guarantee that there will not be any non-null SOC entry. Figure~\ref{fig:curve_seg} illustrates that the charging voltage curves can be split into three linear segments. The length of first segment is approximately ten SOCs. The lengths of the second and third segments vary, which depend on the capacity of the battery. As the FCC decreases, the length of the second segment decreases and the third segment increases. The second segment has a positive slope, whereas the third segment is parallel to X-axis. For simplicity, we consider only the second and third segments.  We use linear interpolation to estimate the missing values in the reference voltage curves (\textit{line 4, Alg.~\ref{alg:cevent}}). 

Neither the 250 samples nor the interpolation guarantees that the voltage for a particular SOC in the reference curve is not an outlier. Therefore, we first find the absolute voltage difference for two consecutive SOCs. This list of differences is a normal distribution and  we next apply iterative $Grubbs~test$ on this distribution to detect the outliers in the voltage curve using the Matlab function presented in~\cite{matout} (\textit{line 5, Alg.~\ref{alg:cevent}}).  Grubbs test determines whether the tested value is the highest/lowest and furthest from the sample mean~\cite{doutlier}. We again apply interpolation to replace the outliers.  
We also find the CC phase length from the reference voltage curve. It is the SOC value when the voltage reaches the maximum 4.2/4.35$\pm$0.05V  (i.e. $mcc~in~ Alg.\ref{alg:cevent}$). In Table~\ref{tab:carat_results}, we present the lengths of CC phases of top 14 models and notice that Galaxy S2/3/4's CC phase lengths are close to what we measured in Section~\ref{sec:three}.

\subsection{Reference C-rate Curves}

\begin{table}[t]
\begin{center}
            \caption{Top fourteen device models, and their battery properties ordered with respect to the number of samples in the descending order. The columns represent the CC phase length in SOC, reference C-rates derived from the crowd, the C-rates computed from the ratio of charging current (mA) and the battery capacity (mAh), users with more than 25 samples within the CC-phase,  and the percentages of the users with lower FCCs.}

  \footnotesize
    \begin{tabular}{|p{18mm}|p{4mm}|p{8mm}|p{10mm}|p{6mm}|p{5mm}|}
      \hline
       \textbf{Device\break Model}  & \textbf{CC} & \textbf{Crowd\break C-rate} & \textbf{C-rate\break$\frac{mA}{mAh}$}& \textbf{Users}& \textbf{Poor\break FCC}\\\hline
       Nexus 4 &84& 0.47& \textit{0.47},$\frac{1000}{2100}$&920&51\%\\\hline 
       SCH-I535&81 &0.39&0.44,$\frac{925}{2100}$&619&75\%\\\hline
       SPH-L710 &79&0.42&0.44,$\frac{925}{2100}$&439&54\%\\\hline
       Nexus 7& 91&0.25&\textit{0.34},$\frac{1350}{3950}$&532&74\%\\\hline
       GT-N7100&84&0.43& \textit{0.52},$\frac{1600}{3100}$&327&49\%\\\hline
       SGH-T999 & 80&0.44&0.44,$\frac{925}{2100}$&284&60\%\\\hline
       SGH-I747M & 80&0.42&0.44,$\frac{925}{2100}$&232&71\%\\\hline
       G. Nexus & 74&0.59&\textit{0.54},$\frac{1000}{1850}$&341&59\%\\\hline
       SCH-I605 & 83&0.41&\textit{0.45},$\frac{1400}{3100}$&167&57\%\\\hline     
       GT-I9100(GS2)&  76&0.35&0.39,$\frac{645}{1650}$&421&66\%\\\hline
       GT-I9505(GS4)& 75&0.63&0.6,$\frac{1560}{2600}$&318&45\%\\\hline
       SCH-I545&  74&0.6&0.6,$\frac{1560}{2600}$&174&48\%\\\hline
       GT-I9300(GS3)& 83&0.47&0.44,$\frac{925}{2100}$&284&58\%\\\hline
       Nexus 5	& 91&0.55& \textit{0.52},$\frac{1200}{2300}$&253&44\%\\\hline
    \end{tabular}

     \label{tab:carat_results}
    \end{center}
\end{table}

Similar to the voltage curve, we also construct model specific  rate curves. To this end, we first select the one percent charging time curve constructed from the reference voltage curve samples in Section~\ref{sub:methodology}. Unlike the voltage curve, the charging time curve is required to have charging time for all the SOCs within the CC phase boundary, which is equivalent to the first two segments of the voltage curve. The line 7 in Alg.~\ref{alg:cevent} shows that the length of the charging time curve is equivalent to the CC phase length determined from the voltage curve.  As shown in Figure~\ref{fig:onePercurve}, these curves are almost parallel to X-axis within the SOC boundary. Therefore, we simply use linear interpolation to predict the missing values. Naturally the corresponding time curve also may contain outliers. The figure shows that the charging time within the SOC boundary follows normal distribution and thus we apply the $Grubbs~test$ and interpolation to find and replace the outliers (\textit{line 8, 9, Alg.\ref{alg:cevent}}).

Once we have the charging time curve of a model, we apply equation \eqref{eq:one} to obtain the cumulative rate curve. From the rate curve we select the C-rate of the CC phase SOC boundary. The third column in Table~\ref{tab:carat_results} presents the C-rate computed from the data set. We find that the Galaxy S2/S3/S4 device rates are within $\pm0.05$C compared to what we measured in Section~\ref{sec:fcc}.  

We further investigate the effectiveness of our crow-sourced rate estimation technique. We looked for the battery capacity and the charging current for some other models in the table. We have identified SCH-I535, SPH-L710, SGH-I747M, SGH-T999 are Galaxy S3 models, and SCH-I545 is Galaxy S4 model. From Internet, we have also collected the capacity and the charging current of other models. These C-rates are of less confidence  and  marked italic in the fourth column of the table. We notice  that the crowdsourced reference rates are within $\pm$0.09 C of their computed values for these popular models. From the model-specific rate we find the coefficients of the capacity model as discussed in Section 4.3.

\subsection{User Rate Curves and FCC in the Wild}
\label{sub:fcc_loss}

We next find the devices with reduced FCC. In order to do that we compare the rate from the model-specific reference curve with the rate from user specific rate curve (\textit{line 12-23, Alg.~\ref{alg:cevent}}). As our interest is the latest battery capacity of a device, we construct the one percent charging time curve from the voltage curve of a device from the samples of the latest month available.  We select the maximum voltage per SOC, as it guarantees less device utilization and the recent state of the battery at the same time. Again, it is important to have adequate number of samples for a user curve as well. We consider only those users who had  at least 25 samples within the SOC boundary of the second segment. 
Similar to the model reference curves, we also detect outliers and apply linear interpolation for estimating the missing values in the user-specific curves. Once, we have the user C-rate, we compute the capacity loss using the model.

Among 9560 user devices, 5311 devices of 333 models had reduced battery capacity. The sixth column in Table~\ref{tab:carat_results} shows that more than 50\% of the devices of nine popular models had reduced battery capacity. The ratio of such users is the  lowest with Galaxy S4 (GT-I9505) and Nexus 5. On the other hand, more than 70\% Galaxy S2, Nexus 7, and SCH-I535 devices suffered from capacity loss. The range of FCC losses for these users are illustrated as boxplots in Figure~\ref{fig:devloss}. We can see that most of the devices of these models had less than 40\% capacity loss. We looked into the length of CC phases of user devices and found that around 38 devices among all the devices had CC phase length of zero. In other words, these devices had significant capacity loss.

\section{Discussion}
\label{sec:seven}

The relation between battery capacity and charging cycles have been studied linear. In this section, we try to foster the relation between battery capacity and the relative charging rate. We also emphasize the limitation of our FCC estimation methods.
\subsection{C-rate Vs. Capacity Modeling}

\begin{figure}[t]
  \begin{center}
\includegraphics[width=1.0\linewidth,height = 0.6\linewidth]{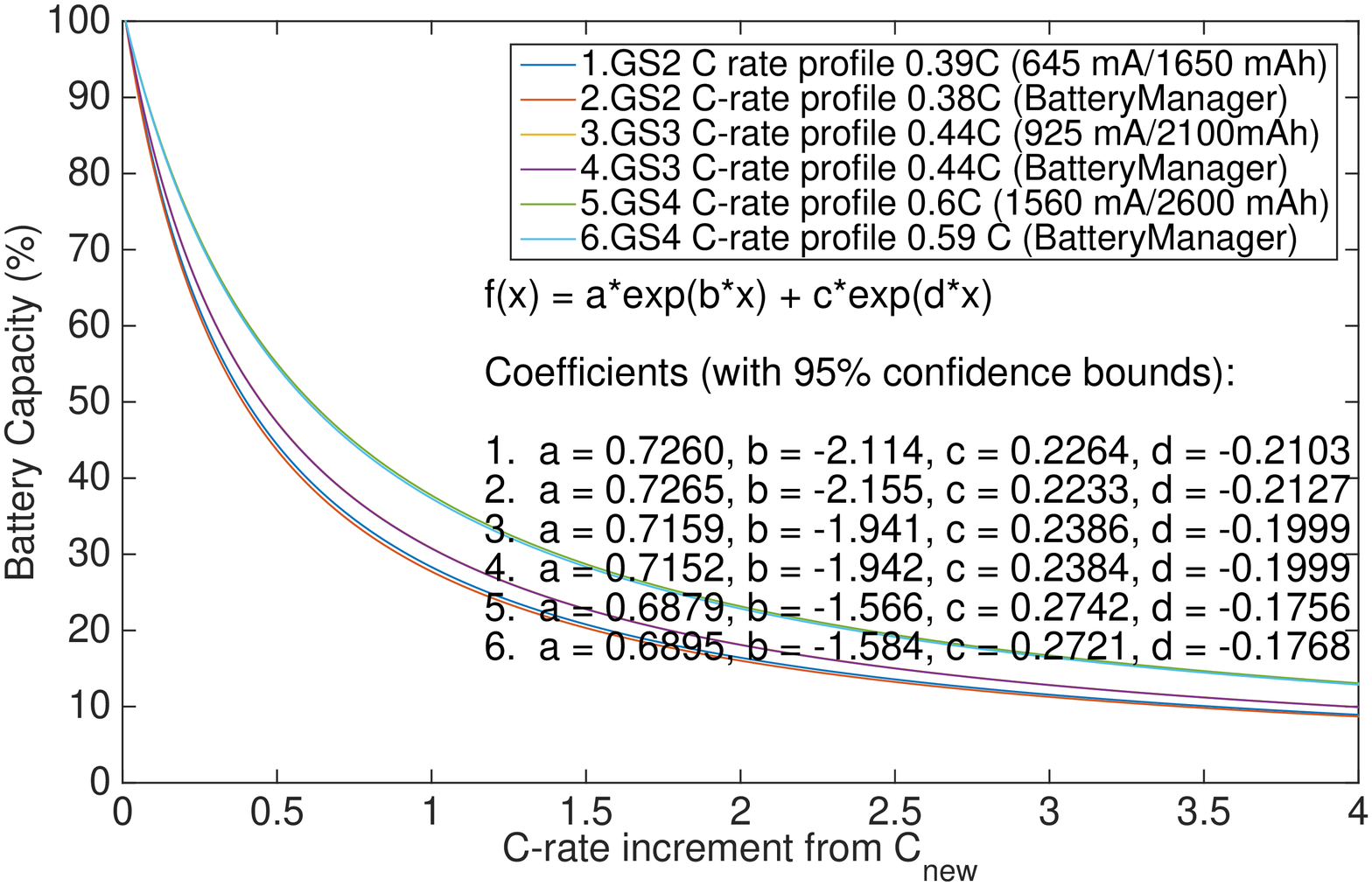}
    
    \caption{{\bf The C-rate based capacity models of the Galaxy devices. } {\sl FCC reduction is the exponential decay function of the rate increment from the $C_{new}$.}}
    \vspace{-5mm}
    \label{fig:crate_profiles}
 \end{center}
 \end{figure}

We look for a general relationship between C-rate and FCC. We have shown that the C-rate increases as the capacity decreases. Now, given the C-rate using equation \eqref{eq:crate2}, we can find the recent FCC of a battery. In the equation, we increase the value of $C_{now}$ by 0.01 C and use the measured charging current. We  find the corresponding capacity and plot in Figure~\ref{fig:crate_profiles}. We notice that for all our experimental devices, the relation between FCC and C-rate is an exponential decay function. From the figure it is also obvious that the decay function is model specific. In this way, it is possible to construct an exponential model or a profile consisting of capacity and rate value pairs for the battery of a device. Once we have the profile, we simply just map the C-rate with the corresponding capacity. In the figure, we can see that capacity reduces almost 90\% when  C-rate increases by 2 C from the initial rate. The models for the experimental devices are annotated in Figure~\ref{fig:crate_profiles}.

\begin{align}
\label{eq:two}
FCC_{now} = a\times e^{b\times (C_{now}-C_{new})} + c\times e^{d\times (C_{now}-C_{new})}
\end{align}

Now, for the C-rate from the BatteryManager updates, it is not straightforward to construct such model or profile, as most of the smartphones do not provide capacity and rate  information. However, this C-rate can be expressed as the ratio of an arbitrary charging current and a capacity, e.g., Galaxy S3 C-rate derived from the BatteryManager updates, $0.44 ~C~=~44/100$. As long as the ratio remains equal, it is possible to construct a similar capacity estimation model. Figure~\ref{fig:crate_profiles} shows  the comparison of the models constructed from two kinds of rates and we notice that the coefficients of the models are close of the respective devices. The goodness of their fits also have similar measures where,  R-square is 0.99 and RMSE is 0.007-0.008 for all devices. Therefore, a general model can be expressed as ~\eqref{eq:two}. If we plug in the difference of $C_{now}$ and $C_{new}$ values from Table 2 to the \eqref{eq:two} with the corresponding device model coefficients, we would get the similar estimates presented in Table.

\subsection{Limitations}
In this work, we have used the charging rates with both the AC and USB2.0 chargers. The proposed FCC estimation technique works regardless of underlying SOC estimation technique or fuel gauge chip is being used by a device, but it's accuracy depends on the performance of these fuel gauge chips. We have shown that their accuracy is higher while charging than discharging. However, if the capacity reduces significantly so that the battery never experiences CC phase, our approach would underestimate the capacity loss. Therefore, our approach is applicable as long as the length of the CC phase is higher than zero. The relation between FCC and the length of CC phase is our ongoing research.

The limitation of crowdsourced approach is the number of unbiased samples and we have mentioned earlier that there are two competing  sources of bias that affect the construction of the reference voltage and rate curves; the samples gathered during active usage of the device and the samples from the lower FCC devices. For the active usage samples, the reference rate of a model would be lower than the measured value in the laboratory and the crowdsourced approach would overestimate the capacity loss. For the latter case, model-specific rate would be higher than the measured value and the capacity loss would be underestimated. It is possible to overcome this bias by learning the rates as new unbiased samples arrive at the Carat back-end from the users. This  method is being integrated as Carat's battery diagnosis feature.

\section{Related Work}
\label{sec:eight}

Significant amount of research work focused on the energy consumption measurement and optimization of different applications and system~\cite{Balasubramanian:2009, Carroll:2010,Miettinen:2010,Pathak:2012,Wang:2009}. A large body of research investigated the energy consumption of mobile devices through profiling, modeling, and debugging~\cite{hoquecsur2015}. The profiling methods  include novel techniques to trace the energy consumption from code to different hardware components \cite{Dong:2011,Pathak:2011,Pathak:2012}. Such profilers also depend on the power consumption modeling. The state-of-the art power modeling approaches rely on on-device resource profiling and battery information. PowerBooter~\cite{powerbooter} relies on SOC updates and OCV discharge curves to build the regression based power models. V-edge~\cite{Xu2013} and BattTracker~\cite{Koo:2016} rely on SOC updates, OCV and the load voltage to model power consumption.  Sesame~\cite{Dong:2011}, AppScope~\cite{Yoon:2012}, and DevScope~\cite{Jung:2012} rely on SOC updates and current drawn estimates from the Coulomb counter-based fuel gauges. Carat relies on SOC updates to compute the energy ratings of different applications for different energy models towards finding the energy hogs and bugs. We also depend on battery manager updates. We specifically use SOC update time to find the charging rates and use the battery voltage to determine the length of the CC phase. However, we estimate the FCC of the battery. Our approach works irrespective the SOC estimation or fuel gauge chip used by the device.

The discharge rate of the battery increases for the same usage~\cite{barr2014}, as the Lithium-Ion battery ages. Consequently, the reported power consumption will be higher than the true value for an aged battery. The accuracy of the self-metering profilers is intertwined with the age of the battery and will decrease as the age of the battery increases. 
For this reason, the self-metering profilers may require retraining of their models. Precise quantification of its effect on the accuracy would require further evaluation of the profilers with batteries of different ages and so far has been an open question. Battery FCC estimation model enables more accurate modeling by identifying the inefficiency of the energy source of the smartphone.

A number of data-driven strategies exist that predict the capacity as the battery ages and the number of charging cycle increases. Yin et al.~\cite{Shan2013} and Liu et al. \cite{Liu2013832} applied a few variations of the Gaussian process regression to predict the capacity as a function of charging cycle. The earlier approaches rely on the vendor provided data on FCC and charging cycles. Barr{\'{e}} et al.~\cite{barr2014} studied how the discharge rates can be used to predict the life of Lithium-Ion battery in electric vehicles. Based on the discharge voltage and current during an activity, such as acceleration, they proposed a data driven real-time battery capacity prediction framework.

Compared to these related work, we focus on estimating capacity proactively. Our technique does not require any additional hardware or system modification. Therefore, the  approach can be easily implemented as a part of the mobile system and can be integrated with different applications or operating system initiated optimizations. For example, the SDB proposed by Badam et al.~\cite{Badam:2015} can include a FCC aware battery scheduling, where SDB will learn the FCC of an individual battery while charging and then schedule the batteries accordingly while discharging.

\section{Conclusions}
\label{sec:nine}

In this article, we have shown that the battery voltage and rate curves can capture the FCC of Lithium-Ion batteries. Based on this observation, we proposed and validated an online mechanism to estimate the FCC or the capacity loss. We also implemented and validated a crowdsourced mechanism. We found 30-57\% of devices of popular models having significant capacity loss in a large data set of mobile devices. Compared to the traditional approaches, our approach is device based and can be used to debug the performance of smartphone batteries. In addition, this work paves the way of modeling and implementing FCC-aware energy optimizations of mobile systems and applications.

\section*{Acknowledgment}
This work was funded by the Academy of Finland CUBIC project with grant numbers 277498 and 278207.


%




\ifCLASSOPTIONcaptionsoff
  \newpage
\fi





%


%
\vspace{-2cm}
\begin{biography}[{\includegraphics[width=1in,height=1.2in,clip,keepaspectratio]{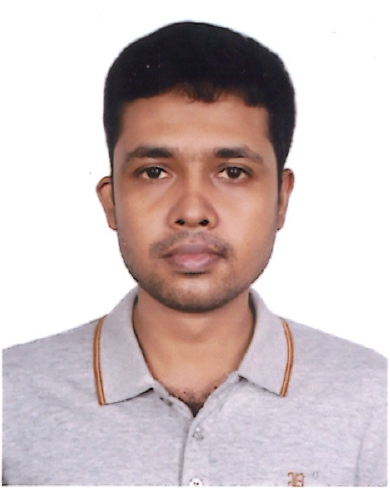}}]{Mohammad Ashraful Hoque}
obtained his M.Sc degree in Computer Science and Engineering in 2010, and Ph.D in 2013 from Aalto University. He is currently a postdoctoral researcher in Helsinki Institute for Information Technology (HIIT) and University of Helsinki. His research interests are energy efficient mobile computing, data analysis, distributed computing, traffic and computing resource scheduling, and machine learning.
\end{biography}
\vspace{-2cm}
\begin{biography}[{\includegraphics[width=1in,height=1.2in,clip,keepaspectratio]{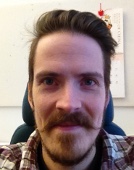}}]{Matti Siekkinen}
obtained the degree of M.Sc. in computer science from Helsinki University of Technology in 2003 and Ph.D from Eurecom / University of Nice Sophia-Antipolis in 2006. He is currently a postdoctoral research fellow at Aalto University. His current research focuses on mobile computing and networking with a special interest in mobile multimedia services.
\end{biography}
\vspace{-2cm}
\begin{biography}[{\includegraphics[width=1in,height=1.2in,clip,keepaspectratio]{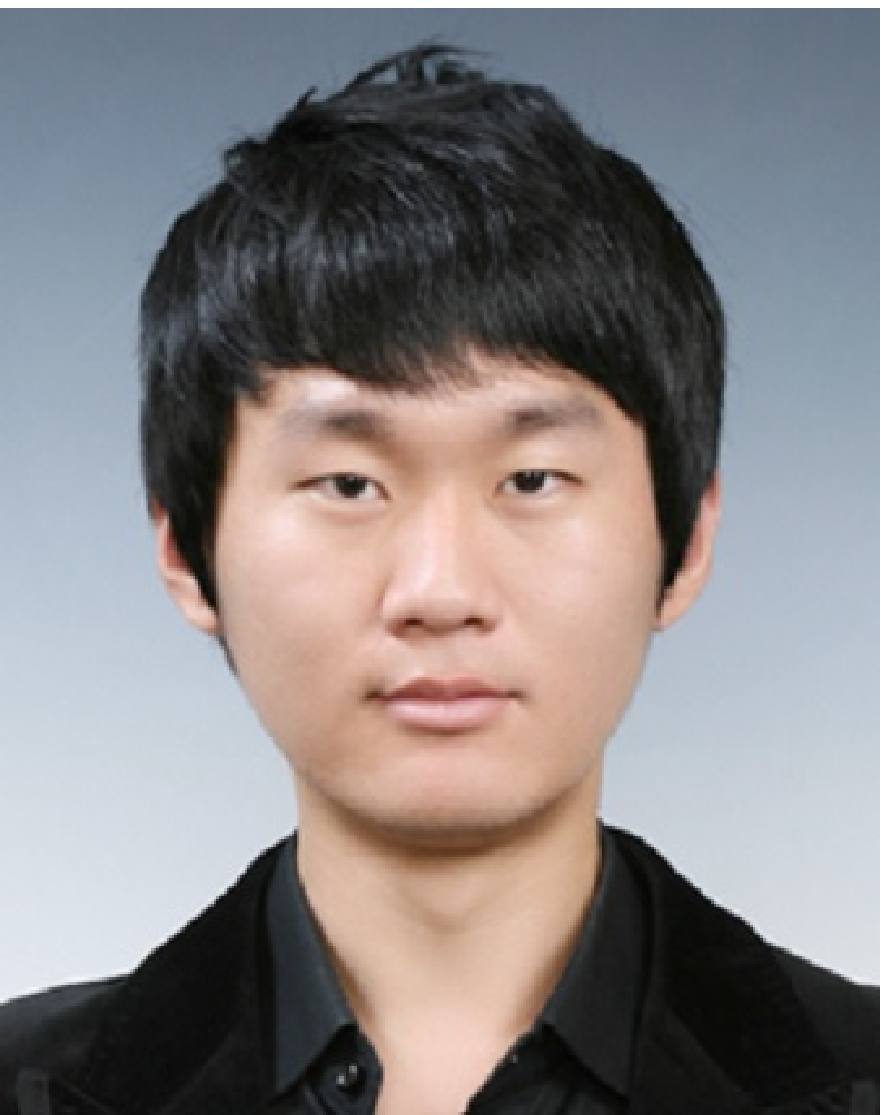}}]{Jonghoe Koo}
(S'12) received the B.S. degree in the Department of Electrical Engineering from Seoul National University, Korea in 2011. 
He is currently working towards a Ph.D. degree in the Department of Electrical and Computer Engineering, Seoul National University, Korea. His current research interests include energy-aware mobile computing and reliable video streaming. He is a student member of IEEE.
\end{biography}

\vspace{-2cm}
\begin{biography}[{\includegraphics[width=1.2in,height=1.2in,clip,keepaspectratio]{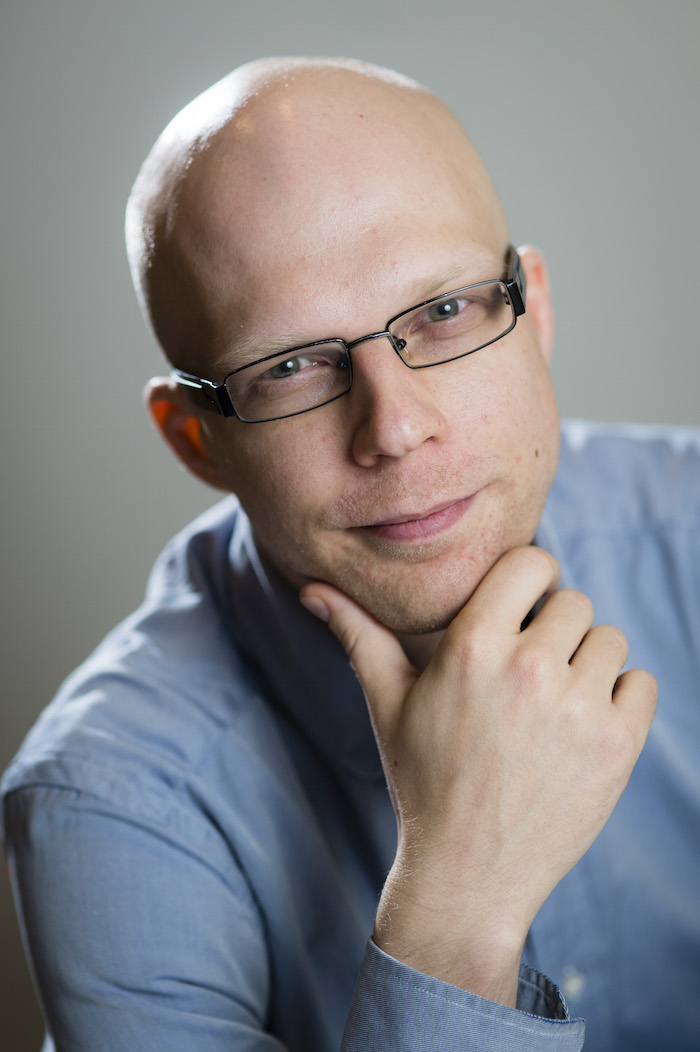}}]{Sasu Tarkoma}
(SMIEEE) is a Professor of Computer Science at the University of Helsinki, and Head of the Department of Computer Science. He has authored 4 textbooks and has published over 160 scientific articles.  His research interests are Internet technology, distributed systems, data analytics, and mobile and ubiquitous computing.  He has seven granted US Patents. His research has received several Best Paper awards and mentions, for example at IEEE PerCom, ACM CCR, and ACM OSR. \end{biography}






\end{document}